\documentclass[12pt,draftcls,onecolumn]{IEEEtran}
\usepackage[utf8]{inputenc}
\usepackage{amsmath,mathtools}
\usepackage{amsthm}
\usepackage{amsfonts}
\usepackage{amssymb}
\usepackage{easyReview} 
\usepackage{soulutf8} 
\DeclarePairedDelimiter\abs{\lvert}{\rvert}
\DeclarePairedDelimiter\norm{\lVert}{\rVert}

\usepackage[justification=centering]{caption}
\usepackage{cite}
\newtheorem{theorem}{Theorem}[]
\newtheorem{lemma}[]{Lemma}
\newtheorem{corollary}{Corollary}
\usepackage{array}
\newcolumntype{C}[1]{>{\centering\let\newline\\\arraybackslash\hspace{0pt}}m{#1}}
\usepackage{algorithm,algorithmic}
\usepackage[list=true]{subcaption}
\usepackage{changebar}
\begin{document}
	\bstctlcite{IEEEexample:BSTcontrol}
	\title{Outage Probability  Analysis of Uplink Cell-Free Massive MIMO Network with and without Pilot Contamination}
	\author{Shashank Shekhar$^{1}$, Muralikrishnan Srinivasan$^{2}$, Sheetal Kalyani$^{1}$, and Mohamed-Slim Alouini$^{3}$ \\
		\thanks{1. Shashank Shekhar and Sheetal Kalyani are with the Dept. of Electrical Engineering, Indian Institute of Technology, Madras, India.
			(Emails: {ee17d022@smail, skalyani@ee}.iitm.ac.in).}
		\thanks{2. Muralikrishnan Srinivasan is with the Dept. of Electrical Engineering, Chalmers University of Technology, Gothenburg, Sweden. (Email: mursri@chalmers.se)} 
		\thanks{3. M.-S. Alouini is with the Computer Electrical and Mathematical Sciences and Engineering (CEMSE) Division, King Abdullah University of Science and Technology (KAUST), Thuwal, Makkah Province, Saudi Arabia. (Email: slim.alouini@kaust.edu.sa) }
	}
	\maketitle
	\begin{abstract}
		This paper derives approximate outage probability (OP) expressions for uplink cell-free massive multiple-input-multiple-output (CF-mMIMO) systems with and without pilot contamination. The system's access points (APs) are considered to have imperfect channel state information (CSI). The signal-to-interference-plus-noise ratio (SINR) of the CF-mMIMO system is approximated via a Log-normal distribution using a two-step moment matching method. OP and ergodic rate expressions are derived with the help of the approximated Log-normal distribution. For the no-pilot contamination scenario, an exact expression is first derived using conditional expectations in terms of a multi-fold integral. Then, a novel dimension reduction method is used to approximate it by the sum of single-variable integrations. Both the approximations derived for the CF-mMIMO systems are also useful for single-cell collocated massive MIMO (mMIMO) systems and lead to closed-form expression. The derived expressions closely match the simulated numerical values for OP and ergodic rate. 
	\end{abstract}
	\section{Introduction}
	Cell-free massive multiple-input multiple-output (CF-mMIMO) system is envisioned as a key enabler for 6G communication systems. \cite{Zhang2019:CFMM_paradigm,akyildiz20206g,zhang2020prospective,matthaiou2021road}. CF-mMIMO system contains many access points (APs) that are connected to a central processing unit (CPU) and jointly serve all the user equipment (UE) by coherent joint transmission and reception \cite{Ngo2015:CellFree_SPAWC, Nayebi2015:CFMMS_Asilomar}. The name cell-free comes from the fact that there are no boundaries, and each AP serves all the UEs. This differs from a conventional small-cell system where each AP serves only a particular set of UEs. Early works in \cite{Ngo2017:CellFree, Nayebi2017:CellFree} compare the CF-mMIMO system with a conventional small-cell system and show a multifold improvement in 95\%-likely throughput can be expected from CF-mMIMO.
	
	\subsection{Prior Art}
	The spectral efficiency (SE) has been studied extensively for both uplink and downlink CF-mMIMO systems for various receiver schemes and fading channels. The authors in \cite{Nayebi2016:CFMMS_Receiver} analyze the SE of the uplink CF-mMIMO system under Rayleigh fading, with minimum mean squared error (MMSE) and large-scale fading deciding (LSFD) receivers.  Upper and lower bound on the SE of the uplink CF-mMIMO system with zero-forcing receiver with perfect and imperfect CSI have been derived in \cite{liu2019spectral}. In \cite{Bashar2019:Uplink_CFMMS}, an achievable rate expression is derived, and then the minimum uplink achievable rate is maximized under per-user power constraint for the CF-mMIMO system. In the downlink scenario, the CF-mMIMO system with APs having multiple antennas was considered in \cite{Nguyen2017:EEinCFMM_ZF, Ngo_2018:CellFreeTotal_Energy}. Here, the system's total energy efficiency was maximized through the power allocation algorithm and AP selection scheme. The uplink and downlink SE of a CF-mMIMO system over Rician fading where the phase shift of the line-of-sight (LoS) component is modeled as a uniform random variable (RV), were analyzed in \cite{Emil2019:RicianTWC}. 
	\par Furthermore, several prior works in literature have comprehensively analyzed the SE for different hardware constraints. For example, the downlink SE of the CF-mMIMO system with low-resolution ADC at APs and UEs is investigated in\cite{hu2019cell}. They considered multiple antennas at APs and a single antenna at UEs. It was found that by increasing the number of antennas at the APs, the performance loss due to low-resolution ADCs at the APs can be mitigated. In \cite{zhang2019performance}, authors consider the CF-mMIMO system with multiple antennas at APs and UEs. The low-resolution ADCs presence was considered only at the APs, and the uplink SE of the considered system was derived. The authors in \cite{Zhang2018:CFMM_HardImp} considered a CF-mMIMO system with transceiver hardware impairment and derived the achievable SE for both uplink and downlink. A CF-mMIMO system with a limited capacity link between APs and CPU \textit{i.e.,} only the quantized signal is assumed to be available at CPU, is considered in \cite{bashar2019max}.
	
	\par The robustness of a CF-mMIMO system in the presence of an active eavesdropper was studied in \cite{Hoang2018:CFSecurityPC}. The authors in \cite{papazafeiropoulos2020performance} derived the coverage probability of the CF-mMIMO system using the tools from stochastic geometry under the assumption that the AP locations follow the Poisson point process.  Some fundamental aspects of the CF-mMIMO system, like channel hardening and favorable propagation, are investigated in \cite{Chen2018:ChanHard_FavProp_CFMM_SG} using stochastic geometry. It was shown that channel hardening is not expected in general, but for the three-slope pathloss model \cite{Ngo2017:CellFree} and multiple antennas at APs, it can be achieved. Similarly, favorable propagation is experienced better under smaller pathloss and higher antenna density. A reliable rate for each user was studied for an asymptotic regime in \cite{koyuncu2018performance}. Finally, there has been a detailed SE analysis of variants of the CF-mMIMO system, known as the user-centric CF-mMIMO system (UC CF-mMIMO) in  \cite{Buzzi2017:CFMM_UC,buzzi2019user,alonzo2019energy,shekhar2022joint}. In a UC CF-mMIMO system, each AP serves only a predefined number of UEs rather than all. Recently, in \cite{nguyen2022spectral}, authors proposed the hybrid relay-reflecting intelligent surface-assisted CF-mMIMO system where the SE for the uplink and downlink of the proposed system was analyzed. Most of the papers mentioned above derived and analyzed SE using the popular use-and-then-forget (UaTF) bound. Also, one can utilize the use-and-then-forget bound only for deriving a lower bound on the SE rather than for other vital metrics, such as outage probability (OP) that depend on the tail characteristics of the signal-to-interference-plus-noise ratio (SINR).

	\subsection{Characterization of Outage Probability (OP)}
	For deriving any expressions for OP, characterization of the probability density function (PDF) or the cumulative distribution function (CDF) of the SINR at the APs is imperative. However, in a CF-mMIMO system, for Rayleigh fading, the numerator and denominator of the SINR are sums of correlated Gamma random variables (RVs). Determining the PDF/CDF of the ratio of correlated Gamma RVs is mathematically intractable \cite{Suman2015:OutageKappa}. Therefore, to the best of our knowledge, there has been no prior literature analyzing the OP of a CF-mMIMO system.
	\par Even in massive MIMO (mMIMO), there have been very few papers studying OP, all of which consider various approximations. For example, in \cite{Feng2016:PoutMassiveMIMO}, the authors consider the OP of a downlink mMIMO system with matched-filter precoding. The numerator term of SINR is treated as a deterministic quantity replaced by its mean, and the interference term is treated as an RV. The PDF of SINR can therefore be obtained by a simple transformation of the interference term's PDF. A similar method is used in \cite{Ding2018:OutageADCMassiveMIMO}, where the base station (BS) is equipped with ADCs of different resolution levels. Here, it is shown that the squared coefficient of variation (SCV) of all the terms except one of the interference terms approaches zero as the number of antennas approaches infinity. Therefore, one can determine the \emph{approximate} PDF of the SINR by determining the PDF of that interference term. However, it may not always be possible to show that the SCV of all but one of the terms of the SINR becomes zero. Even if two non-zero terms exist, the PDF becomes intractable to characterize, and such is the case in CF-mMIMO. In \cite{srinivasan2019analysis}, the SINR is approximated to a Gamma RV by moment matching. The OP is then obtained in terms of the CDF of the Gamma RV. However, the efficacy of moment-matching depends on the distribution to which the metric is matched. Also, in many cases, it is algebraically complex to determine the expressions for the moments\footnote{We conducted trials to approximate the SINR RV by Gamma RVs. However, the resultant expressions did not match the simulated OP. This is elaborated in  Section \ref{Sec: CFOP_Outage_Results}.}. In a few other works, such as \cite{Atapattu2017:ExactOutageMIMO, Beiranvand2018:MRCMassiveMIMO}, the exact expressions for OP are derived under perfect CSI and i.i.d. channel.
	
	\subsection{Contributions}
	For CF-mMIMO, assuming perfect CSI knowledge at the APs and i.i.d. channels are not practical \cite{Ngo2017:CellFree}. Also, it is essential to consider the effect of pilot contamination during the channel estimation phase on the resultant OP. Although an exact expression for OP can be written using conditional expectation, which results in a multi-fold integral of order $M$, where $M$ denotes the number of APs. It is generally challenging to evaluate such multi-fold integrals even numerically when $M$ is large, which is the typical case for the CF-mMIMO systems. Therefore, this paper uses two approaches to obtain novel OP expressions for the CF-mMIMO system. For the case of no-pilot contamination, to evaluate the multi-fold integrals, we propose to exploit a uni-variate dimension reduction method to reduce the $M$th order integration into $M$ single-order integration approximations \cite{Rahman2004:Integral_DimensionReduction}. Secondly, for the case of pilot contamination, we provide a two-step moment matching method to approximate the SINR by a Log-normal RV from which OP can be directly evaluated. We compare and contrast our work to the existing critical literature in Table \ref{Tab: CFOP_prior_work}. The obtained expressions are straightforward to evaluate and novel. Our contributions through this paper are summarized as follows:
	\begin{itemize}
		\item \textbf{Two-step moment matching method} We study the OP of the uplink of the CF-mMIMO system with and without pilot contamination for Rayleigh fading. We approximate the SINR with a Log-normal RV using a two-step moment matching method. A simple approximation for the OP is obtained using the Log-normal CDF. 
		\item \textbf{Uni-variate dimension reduction method:} For the case of no-pilot contamination, we derive an exact expression for the OP in terms of a multi-fold integral. A novel approximation is then derived by reducing the multi-fold integral using the uni-variate dimension reduction method\cite{Rahman2004:Integral_DimensionReduction}.
		\item \textbf{Special cases:} Using the SINR as mentioned earlier characterization, we propose an alternative to UaTF for evaluating SE. Furthermore, we obtain approximate OP expressions in terms of simple elementary functions for a single-cell collocated mMIMO system. 
	\end{itemize}
	
	\begin{table}[]
		\centering
		\begin{tabular}{|C{1.5cm}|C{1.5cm}|C{1cm}|C{4cm}|C{1.5cm}|}
			\hline
			Reference & Scenario  & Metric & Methodology & Imperfect CSI \\
			\hline
			\cite{Ngo2017:CellFree,Nayebi2017:CellFree} & Cell-Free & SE & Use and forget bound & $\checkmark$ \\
			\hline 
			\cite{Feng2016:PoutMassiveMIMO} & mMIMO & OP  & SCV & \texttimes \\
			\hline
			\cite{Ding2018:OutageADCMassiveMIMO} & mMIMO & OP & SCV & $\checkmark$ \\
			\hline
			\cite{srinivasan2019analysis} & mMIMO & OP & Moment matching & \texttimes \\
			\hline
			\cite{Atapattu2017:ExactOutageMIMO} & mMIMO & OP & Exact analysis & \texttimes \\
			\hline
			\cite{Beiranvand2018:MRCMassiveMIMO} & mMIMO & OP & Moment matching & \texttimes \\
			\hline
			This work & Cell-Free \& mMIMO & OP & Two-step moment matching \& uni-variate dimension reduction & $\checkmark$  \\
			\hline
		\end{tabular}
		\caption{Our paper vis-a-vis key existing literature}
		\label{Tab: CFOP_prior_work}
	\end{table}
	
	\subsubsection*{Organization} The rest of the paper is structured as follows. In Section \ref{Sec: CFOP_SystemModel}, the considered system model of the CF-mMIMO system is discussed in detail. Section \ref{Sec: CFOP_Outage_OPAnalysis} presents the analytical expression of OP and rate obtained via two-step moment matching and uni-variate dimension-reduction approach. The simulation results and discussion is given in Section \ref{Sec: CFOP_Outage_Results} and finally the conclusion are drawn in Section \ref{Sec: CFOP_Conclusion}   
	\subsubsection*{Notation} In this paper, $\mathcal{CN}\left(a,b\right)$ denotes the complex Gaussian distribution with mean $a$ and variance $b$. $\operatorname{LN}(\mu, \sigma^2)$ represent the Log-normal distribution with parameters $\mu$ and $\sigma$. The mean and variance of RV $ X $ are denoted by $\mathbb{E}\left[X \right]$ and $ \mathbb{V}\left[X \right] $. $\operatorname{Cov}\left( X,Y\right)$ represents the covariance between RVs $X$ and $Y$. $\operatorname{diag}(a_{1},\cdots,a_{M})$ denotes a $M \times M$ diagonal matrix with entries $a_{1},\cdots,a_{M}$, and $\mathbf{I}_{N}$ represents the identity matrix of size $N$. Also,  $ \left(a\right)_{n} $ denotes the Pochhammer symbol,  $ \operatorname{U}\left(\cdot \right) $ is the unit step function, and $\mathfrak{Re}\left(z \right)$ is the real part of $z$.
	
	\section{System Model}\label{Sec: CFOP_SystemModel}
	We consider a cell-free massive MIMO system with $ M $ APs and $ K $ users where $ M \gg K $, \textit{i.e.}, the number of APs is much more than that of users. Each AP is equipped with $ N $ antennas, and the users are equipped with a single antenna. The channel between the $ m $th AP and the $ k $th user is modeled as a Rayleigh fading channel.  Let  $ \mathbf{g}_{mk} \in \mathbb{C}^N$, represent the channel vector  between the $ m $th AP and the $ k $th user and we have,
	\begin{equation}
		\begin{aligned}
			\mathbf{g}_{mk} \sim \mathcal{CN}\left(\mathbf{0},\beta_{mk}\mathbf{I}_{N}\right),
		\end{aligned}
	\end{equation}
	where $ \beta_{mk}$  represents the large scale fading coefficients between the $ m $th AP and the $ k $th user. Note that this model is similar to the one assumed in \cite{Ngo_2018:CellFreeTotal_Energy}. We assume that the knowledge of $ \beta_{mk}$ is available at both the AP and the user. Let $ \tau_{c} $ be the length of the coherence interval (in samples). Typically, in a cell-free massive MIMO system, the coherence interval is partitioned into three phases, namely the uplink training phase, uplink data transmission phase, and downlink data transmission phase \cite{Ngo2017:CellFree}. In this work, we do not focus on downlink data transmission. Let $ \tau_{p} $ be the length of the uplink training duration (in samples). Therefore, $\left(\tau_{c} - \tau_{p}\right) $ is the duration of the uplink data transmission phase. The process of uplink training and uplink data transmission is described in the following subsections.
	
	\subsection{Uplink Training}\label{SubSec:CFOP_Outage_UT}
	Before the transmission of uplink data by users, APs will acquire the channel state information (CSI) through a training phase. This acquired CSI is then used to process the received data symbols during the uplink data transmission phase. During this phase, all the users simultaneously transmit their pilot sequence to the APs. Let $\sqrt{\tau_{p}} \boldsymbol{\phi}_{k}  \in \mathbb{C}^{\tau_{p} \times 1}$ be the pilot sequence transmitted by the $ k $th user, $ \forall k = 1, \dots , K $ where $ \Vert \boldsymbol{\phi}_{k} \Vert^{2} = 1 $. The signal received at $ m $th AP during the training phase is
	\begin{equation}
		\begin{aligned}
			\mathbf{Y}_{p,m} = \sqrt{\tau_{p} \rho_{p}} \sum_{k=1}^{K} \mathbf{g}_{mk}\boldsymbol{\phi}_{k}^{H}  + \mathbf{W}_{p,m},
		\end{aligned} 
	\end{equation}
	where $ \rho_{p} $ is the normalized transmit SNR of each pilot symbol, and $ \mathbf{W}_{p,m} \in \mathbb{C}^{N \times \tau_{p}}$ is the noise matrix whose entries are i.i.d. zero-mean complex Gaussian with variance $1$. Now, to estimate the channel coefficient using the observation $ \mathbf{Y}_{p,m}$, we first project the received signal on $ \boldsymbol{\phi}_{k}^{H} $ and then use the minimum-mean squared error (MMSE) estimator. Let $ \check{\mathbf{y}}_{p,mk} \triangleq \mathbf{Y}_{p,m} \boldsymbol{\phi}_{k} $, \textit{i.e.},
	\begin{equation}
		\begin{aligned}
			\check{\mathbf{y}}_{p,mk} 
			&= \sqrt{\tau_{p} \rho_{p}} \mathbf{g}_{mk} 
			+ \sqrt{\tau_{p} \rho_{p}}\sum_{ i \neq k}^{K} \mathbf{g}_{mi}\boldsymbol{\phi}_{i}^{H} \boldsymbol{\phi}_{k} 
			+ \tilde{\mathbf{w}}_{p,mk},
		\end{aligned}
	\end{equation}
	where $ \tilde{\mathbf{w}}_{p,mk} =  \mathbf{W}_{p,m} \boldsymbol{\phi}_{k} $ is a vector with i.i.d. $\mathcal{CN}\left(0,1 \right)$ entries. The MMSE estimator is hence given by,
	\begin{equation}\label{Eq:CFOP_MMSE_g}
		\begin{aligned}
			\hat{\mathbf{g}}_{mk} &= \mathbb{E}\lbrace \mathbf{g}_{mk}\check{\mathbf{y}}_{p,mk}^{H} \rbrace \left(\mathbb{E}\lbrace \check{\mathbf{y}}_{p,mk}\check{\mathbf{y}}_{p,mk}^{H} \rbrace\right)^{-1}\check{\mathbf{y}}_{p,mk},\\
			&= c_{mk} \check{\mathbf{y}}_{p,mk},
		\end{aligned}
	\end{equation}
	where $ c_{mk} = \frac{\sqrt{\tau_{p} \rho_{p}}\beta_{mk}}{\tau_{p} \rho_{p}{\sum\limits_{i = 1}^{K} {\beta}_{mi} \abs{\boldsymbol{\phi}_{k}^{H} \boldsymbol{\phi}_{i}}^{2}} + 1 }. $ Here, the term $ \tau_{p} \rho_{p}{\sum\limits_{i \ne k}^{K} {\beta}_{mi} \abs{\boldsymbol{\phi}_{k}^{H} \boldsymbol{\phi}_{i}}^{2}} $ corresponds to pilot contamination due to the non-orthogonality of the pilot sequence of different users.   Also, $ \hat{\mathbf{g}}_{mk} \sim \mathcal{CN}\left( \mathbf{0},\gamma_{mk} \mathbf{I}_{N} \right)$, where $ \gamma_{mk} = \sqrt{\tau_{p} \rho_{p}}\beta_{mk}c_{mk}.  $ In the case of orthogonal pilots, \textit{i.e.}, no pilot contamination, we have  $ c_{mk} = \frac{\sqrt{\tau_{p} \rho_{p}}\beta_{mk}}{{\tau_{p} \rho_{p}} {\beta}_{mk}  + 1 } $, and therefore, $ \gamma_{mk} = \frac{\tau_{p} \rho_{p}\beta_{mk}^2}{{\tau_{p} \rho_{p}} {\beta}_{mk}  + 1 }$.
	\subsection{Uplink Data Transmission}\label{SubSec:CFOP_Outage_UplinkDT}
	In the uplink data transmission phase, each user transmits its data symbol to all the APs. Let $ p_{k} $ be the symbol of the $ k $th user, such that $ \mathbb{E} \left[ \abs*{p_{k}}^{2} \right] = 1 $.  Hence, the received signal at the $ m $th AP is
	\begin{equation}\label{Eq:CFOP_yum}
		\mathbf{y}_{u,m} = \sqrt{\rho_{u}}\sum_{k=1}^{K} \mathbf{g}_{mk}p_{k} + \mathbf{w}_{u,m},
	\end{equation}
	where $ \rho_{u} $ is normalized uplink SNR and $ \mathbf{w}_{u,m} $ is the additive Gaussian noise with $ \mathbf{w}_{u,m} \sim \mathcal{CN}\left(\mathbf{0}, \mathbf{I}_{N} \right)$. Since all the APs employ MRC, they multiply their copies of the received signal with the estimated channel coefficients $\hat{\mathbf{g}}_{mk}^{H}$. The APs then send their received signal to the CPU\footnote{It is assumed that the APs are connected with the CPU through a flawless backhaul network.}. Therefore, the received signal at the CPU is given by
	\begin{equation}\label{Eq:CFOP_ruk}
		\begin{aligned}
			r_{u,k} &= \sum_{m=1}^{M} \hat{\mathbf{g}}_{mk}^{H} \mathbf{y}_{u,m}, \\
			&=  \sqrt{\rho_{u}}\sum_{m=1}^{M} \hat{\mathbf{g}}_{mk}^{H} \hat{\mathbf{g}}_{mk}p_{k} + \sqrt{\rho_{u}}\sum_{m=1}^{M}\hat{\mathbf{g}}_{mk}^{H} \boldsymbol{\varepsilon}_{mk}p_{k} +\sqrt{\rho_{u}}\sum_{i\ne k}^{K}\sum_{m=1}^{M}\hat{\mathbf{g}}_{mk}^{H} \mathbf{g}_{mi}p_{i} + \sum_{m=1}^{M} \hat{\mathbf{g}}_{mk}^{H} \mathbf{w}_{u,m},
		\end{aligned}
	\end{equation}
	where $ \boldsymbol{\varepsilon}_{mk} $ is the channel estimation error, \textit{i.e.}, $ \mathbf{g}_{mk} - \hat{\mathbf{g}}_{mk} $. $ \boldsymbol{\varepsilon}_{mk} $ is a  $\mathcal{CN}\left(0,\left(\beta_{mk}- \gamma_{mk}\right)\mathbf{I}_{N} \right)$ random vector. The symbol $ p_{k} $ transmitted by the $k$th user is detected using $ r_{u,k} $.  Let $  \hat{\mathbf{g}}_{k} = \begin{bmatrix}
		\hat{\mathbf{g}}_{1k} \ \dots \ \hat{\mathbf{g}}_{Mk} 
	\end{bmatrix}^{T} $ denote the channel estimate vector for $ k $th user  $ \forall k=1,\dots, K$. Note that, $ \hat{\mathbf{g}}_{k} \sim \mathcal{CN}\left(\mathbf{0},\mathbf{C}_{\hat{\mathbf{g}}_{k},\hat{\mathbf{g}}_{k}} \right) $ is $ MN \times 1 $ complex Gaussian vector, where
	\begin{equation}
		\begin{aligned}
			\mathbf{C}_{\hat{\mathbf{g}}_{k} ,\hat{\mathbf{g}}_{k}} = \operatorname{diag}\left(\gamma_{1k} \mathbf{I}_{N},\dots,\gamma_{Mk} \mathbf{I}_{N} \right)
		\end{aligned}
	\end{equation}
	is the covariance matrix of $ \hat{\mathbf{g}}_{k} $. Using (\ref{Eq:CFOP_ruk}), the  effective  SINR of the $ k $-th user is as follows:
	\begin{equation}\label{Eq: CFOP_SINRInner}
		\begin{aligned}
			\lambda_{u,k} = \frac{X_{u,k}}{Y_{u,k}} 
			= \frac{\rho_{u} \norm*{\hat{\mathbf{g}}_{k}}^{4} }{ \rho_{u} \sum\limits_{i\ne k}^{K} \abs*{\mathbf{\hat{g}}_{k}^{H} \mathbf{\hat{g}}_{i}}^{2}  +   \hat{\mathbf{g}}_{k}^{H}\left( \rho_{u}\sum\limits_{i=1}^{K} \boldsymbol{\Lambda}_{i} + \mathbf{I}_{MN} \right)\hat{\mathbf{g}}_{k} },
		\end{aligned} 
	\end{equation}
	where $ \boldsymbol{\Lambda}_{i} = \operatorname{diag}\left( \left(\beta_{1i} - \gamma_{1i} \right) \mathbf{I}_{N},\dots,\left(\beta_{Mi} - \gamma_{Mi} \right) \mathbf{I}_{N} \right) $ is a $MN \times MN$ diagonal matrix, $X_{u,k} = \rho_{u} \left( \norm*{\hat{\mathbf{g}}_{k}}^{2} \right)^{2} $ is the desired signal power over estimated channel, and $Y_{u,k} = \rho_{u} \sum\limits_{i\ne k}^{K} \abs*{\mathbf{\hat{g}}_{k}^{H} \mathbf{\hat{g}}_{i}}^{2}  +   \hat{\mathbf{g}}_{k}^{H}\left( \rho_{u}\sum\limits_{i=1}^{K} \boldsymbol{\Lambda}_{i} + \mathbf{I}_{MN} \right)\hat{\mathbf{g}}_{k}  $ is the interference plus noise power. Note that this SINR expression is similar to the SINR expression given in \cite[Eq. 12]{Bjornson2020:CompetitiveCellFree}. 
	\par Using the effective SINR in (\ref{Eq: CFOP_SINRInner}), one can calculate various performance metrics such as achievable rate, outage probability, etc. In the following section, we derive novel OP approximations utilizing the two-stage Log-normal moment matching and uni-variate dimension reduction method.
	
	\section{Outage Probability Analysis}\label{Sec: CFOP_Outage_OPAnalysis}
	The exact expression for OP involves characterizing the CDF of the SINR at the APs. Note that the numerator and denominator of the SINR involve correlated Gamma RV, and determining the CDF of their ratio is mathematically intractable \cite{Suman2015:OutageKappa}. Exact expressions are tractable only for perfect CSI conditions and i.i.d. channels as in the case of mMIMO channels \cite{Atapattu2017:ExactOutageMIMO, Beiranvand2018:MRCMassiveMIMO}. However, to assume that all the channels from the UEs to APs are i.i.d. or that perfect CSI is known at APs is impractical. In literature, it is common to approximate end-to-end SINR via Gamma or Log-normal RV using the technique of moment matching. This method has been successfully employed for various scenarios like intelligent reflecting surface (IRS) assisted communication system \cite{charishma2021outage}, mMIMO system\cite{srinivasan2019analysis}. The challenging part in such approximation is to derive the moments of SINR $ \lambda_{u,k} $, which becomes more difficult due to the correlation between  numerator $ X_{u,k} $ and denominator $ Y_{u,k} $. To circumvent this issue, a bi-variate Taylor's series-based approximation for the first two moments of SINR is presented in \cite{srinivasan2019analysis}. We tried to mimic the approach described in \cite{srinivasan2019analysis}, but the resultant expressions did not match the simulated OP. This is elaborated in section \ref{Sec: CFOP_Outage_Results}. 
	Through extensive simulation, we discovered that the numerator and denominator are closely approximated via Log-normal RV, separately. Therefore, to derive approximate OP, we approximated the numerator and denominator of (\ref{Eq: CFOP_SINRInner}) as Log-normal RV using moment matching then it is easy to show that ratio, \textit{i.e.,} $ \lambda_{u,k} $ is also a Log-normal RV. The result associated with Log-normal approximation is presented in Theorem \ref{Thm: CFOP_SINRLogNormal} and \ref{Thm: CFOP_RateLogNormal}, which is valid for both scenarios \textit{i.e.,} the system with and without pilot contamination. 
	\begin{theorem}\label{Thm: CFOP_SINRLogNormal}
		For a threshold $ T $, the OP of $ k $th user is approximated as 
		\begin{equation}\label{Eq: CFOP_POutLogNormalPC}
			\begin{aligned}
				P_{out}^{k}\left(T\right) = \mathbb{P}\left(\lambda_{u,k} < T \right) &\approx \frac{1}{2} \operatorname{erfc}\left(-\frac{\ln T - \mu_{\lambda_{u,k}}}{\sigma_{\lambda_{u,k}} \sqrt{2}}\right),
			\end{aligned}
		\end{equation}
		where parameter $ \mu_{\lambda_{u,k}}, \sigma_{\lambda_{u,k}} $ of Log-normal distribution are evaluated as
		\begin{equation}\label{Eq: CFOP_MuLogNormalPC}
			\begin{aligned}
				\mu_{\lambda_{u,k}} &= \mu_{X_{u,k}} - \mu_{Y_{u,k}},
			\end{aligned}
		\end{equation}
		\begin{equation}\label{Eq: CFOP_SigmaLogNormalPC}
			\begin{aligned}
				\sigma _{\lambda_{u,k}} &= \sqrt{\sigma^{2}_{X_{u,k}} +  \sigma^{2}_{Y_{u,k}} - 2 \log\left( \frac{\mathbb{E}\left[X_{u,k}Y_{u,k}\right]}{\mathbb{E}\left[X_{u,k}\right]\mathbb{E}\left[Y_{u,k}\right]}\right)}.
			\end{aligned}
		\end{equation}
		Here, $ \operatorname{erfc}\left(\cdot\right) $ is the complementary error function \cite{erfcdef} and  $ \mu_{X_{u,k}},$ $ \mu_{Y_{u,k}},$ $ \sigma_{X_{u,k}} ,$ $ \sigma_{Y_{u,k}} $ is evaluated using  (\ref{Eq:CFOP_XMuSigma}) and (\ref{Eq:CFOP_YMuSigma}). 
	\end{theorem}	
	\begin{proof}
		Please refer to Appendix \ref{App: CFOP_ProofSINRLogNormal} for the proof.
	\end{proof}   
	\begin{corollary}\label{Cor: CFOP_LogNormalNPC}
		In the absence of pilot contamination, the OP of $ k $th user is approximated using \eqref{Eq: CFOP_POutLogNormalPC}, where \eqref{Eq: CFOP_MuLogNormalPC} and \eqref{Eq: CFOP_SigmaLogNormalPC} are evaluated using the following expression for the moments of $ Y_{u,k} $. 
		\begin{equation}\label{Eq: CFOP_YMeanNPC}
			\begin{aligned}
				\mathbb{E}\left[Y_{u,k}\right] &= N\rho_{u} \sum\limits_{i\ne k}^{K} \sum\limits_{m=1}^{M}\gamma_{mk}\gamma_{mi} + N\sum\limits_{m=1}^{M}\gamma_{mk} +  N\rho_{u}\sum\limits_{m=1}^{M} \sum\limits_{i=1}^{K}\left(\beta_{mi} - \gamma_{mi}\right)\gamma_{mk}.
			\end{aligned}
		\end{equation} 
		\begin{equation}\label{Eq: CFOP_Y2MeanNPC}
			\begin{aligned}
				&\mathbb{E}\left[Y_{u,k}^{2}\right] =  \rho_{u}^{2}\left(\sum_{i\ne k}^{K}N\sum_{m=1}^{M}\gamma_{mk}^{2}\gamma_{mi}^{2} +  \sum_{i\ne k}^{K} \left(N\sum\limits_{m=1}^{M}\gamma_{mk}\gamma_{mi}\right)^{2}+ N\sum_{m=1}^{M}\gamma_{mk}^{2} \left(\sum_{i\ne k}^{K}\gamma_{mi}\right)^{2} \right. \\&
				+ \left. \left(N\sum_{i\ne k}^{K}\sum\limits_{m=1}^{M}\gamma_{mk}\gamma_{mi}\right)^{2} \right)  + N\sum_{m = 1}^{M} \gamma_{mk}^{2} + N^{2}\left(\sum_{m = 1}^{M} \gamma_{mk}\right)^{2} 
				\\&+ \rho_{u}^{2} \left( N\sum_{m = 1}^{M}\sum\limits_{i=1}^{K}  \left(\beta_{mi} - \gamma_{mi}\right)^{2}\gamma_{mk}^{2} + N^{2}\left(\sum\limits_{m=1}^{M}\sum\limits_{i=1}^{K}  \left(\beta_{mi} - \gamma_{mi}\right) \gamma_{mk}\right)^{2}\right) 
				\\&+ 2\rho_{u}\sum_{i\ne k}^{K} \left(N\sum_{m = 1}^{M}\gamma_{mk}^{2}\gamma_{mi} 
				+ N^{2}\left( \sum\limits_{m=1}^{M}\gamma_{mk} \right)\left( \sum\limits_{m=1}^{M}\gamma_{mk}\gamma_{mi}\right) \right) 
				\\&+ 2\rho_{u}\left(N\sum_{m = 1}^{M}\sum\limits_{i=1}^{K}  \left(\beta_{mi} - \gamma_{mi}\right)\gamma_{mk}^{2} +N^{2}\left( \sum\limits_{m=1}^{M}\gamma_{mk} \right)\left(\sum\limits_{m=1}^{M}\sum\limits_{i=1}^{K}  \left(\beta_{mi} - \gamma_{mi}\right) \gamma_{mk}\right) \right)
				\\&+ 2\rho_{u}^{2}\sum_{i\ne k}^{K} \left(N\sum_{m = 1}^{M}\sum\limits_{j=1}^{K}  \left(\beta_{mj} - \gamma_{mj}\right)\gamma_{mk}^{2}\gamma_{mi} + N^{2}\left(\sum\limits_{m=1}^{M}\sum\limits_{j=1}^{K}  \left(\beta_{mj} - \gamma_{mj}\right) \gamma_{mk}\right)\left( \sum\limits_{m=1}^{M}\gamma_{mk}\gamma_{mi}\right) \right).
			\end{aligned}
		\end{equation}
	\end{corollary}
	\begin{proof}
		Equation \eqref{Eq: CFOP_YMeanNPC} and \eqref{Eq: CFOP_Y2MeanNPC} are obtained from \eqref{Eq:CFOP_YMean} and \eqref{Eq:CFOP_Y2} by considering the fact that $ \nu_{mi}^{j} = 0 $ if $ i \ne j $ and $ \nu_{mi}^{i} = \gamma_{mi} $. 
	\end{proof}
	\par Next, we used the Log-normal approximation of $ \lambda_{u,k} $ to derive the approximate ergodic rate of $k$th user. We also derive simple closed-form lower and upper bound on the derived ergodic rate. The results are presented in the following theorem. 
	\begin{theorem}\label{Thm: CFOP_RateLogNormal}
		Given $ \lambda_{u,k} \sim \operatorname{LN}\left(\mu_{\lambda_{u,k}},\sigma_{\lambda_{u,k}}^{2}\right) $ and ergodic rate of $ k $-th user,  is 
		\begin{equation}\label{Eq: CFOP_ER_PC}
			\begin{aligned}
				R_{u,k}  =  \mathbb{E}\left[\log_{2}\left(1 + \lambda_{u,k}\right)\right]  &\approx \frac{1}{2} \int_{0}^{\infty} \operatorname{erfc}\left(\frac{\ln\left(2^{t} - 1\right) - \mu_{\lambda_{u,k}}}{\sigma_{\lambda_{u,k}} \sqrt{2}}\right) dt. 
			\end{aligned}
		\end{equation} 
		\begin{equation}\label{Eq: CFOP_ERBs_PC}
			\begin{aligned}
				\log_{2}\left(\operatorname{e}^{\mu_{\lambda_{u,k}}} + 1\right) < &R_{u,k} < \log_{2}\left(\operatorname{e}^{\mu_{\lambda_{u,k}}} + 1\right) + \frac{\operatorname{e}^{-\mu_{\lambda_{u,k}}}}{\ln 2\left(1 + \operatorname{e}^{-2\mu_{\lambda_{u,k}}}\right) }\left(\operatorname{e}^{\frac{\sigma_{\lambda_{u,k}}^{2}}{2}} - 1\right).
			\end{aligned}
		\end{equation}
	\end{theorem}
	\begin{proof}
		Please refer to Appendix \ref{App: CFOP_ProofRateLogNormal} for the proof.
	\end{proof}
	\begin{corollary}\label{Cor: CFOP_NPC_Rate_LogNormal}
		In the case of no pilot contamination also, the ergodic rate and the respective lower and upper bounds of $ k $-th user are computed using the \eqref{Eq: CFOP_ER_PC}, and  \eqref{Eq: CFOP_ERBs_PC} with the parameters of Log-normal are calculated using \eqref{Eq: CFOP_YMeanNPC} and \eqref{Eq: CFOP_Y2MeanNPC}.
	\end{corollary}
	Along with this, for the case of no pilot contamination, we first derive the conditional OP assuming that $ \hat{\mathbf{g}}_{k}$ is a constant. We then integrate the conditional OP over $\hat{\mathbf{g}}_{k}$, which gives an exact expression of OP in terms of a multi-fold integral of the order $MN$ that is difficult to be solved in close form or evaluated in Mathematica/MATLAB for usual values of $M$ and $N$. Therefore, we explore the use of a dimension-reduction method known as a uni-variate approximation that approximates $MN$th order integration with $MN$ single order integrals. The result associated with the uni-variate approximation is given in the Lemma \ref{Lem: CFOP_OPUnivariate} and Theorem \ref{Thm: CFOP_OPUnivariate}.
	\begin{lemma}\label{Lem: CFOP_OPUnivariate}
		In the absence of pilot contamination, the OP of $k$th user for threshold $ T $ is given as
		\begin{equation}\label{Eq: CFOP_PoutOrthoSim2}
			\begin{aligned}
				P_{out}^{k} \left(T\right) &= 1 - \int\dots\int  \left( \sum_{\substack{i=1 \\ i \ne k}  }^{K}  \frac{\theta_{i}^{K-2}}{\prod\limits^{K-1}_{{\substack{j=1 \\ j \ne i}}} \left(\theta_{i} - \theta_{j} \right)} \left[ 1 -e^{-\frac{\delta_{k}^{T}}{\theta_{i}}}  \right] \right)\operatorname{U}\left(\delta_{k}^{T}\right)  \prod\limits_{m=1}^{M} \prod\limits_{n=1}^{N}e^{-x_{mn}} d \mathbf{x},
			\end{aligned}
		\end{equation}
		where $\mathbf{x} = \left[ x_{11},\dots,x_{1N},x_{21},\dots,x_{MN} \right]$, $ \theta_{i} = \sum\limits_{m=1}^{M} \left(\sum\limits_{n=1}^{N} x_{mn}  \right)\gamma_{mk} \gamma_{mi} $, and 
		
		\begin{equation}
			\begin{aligned}
				\delta_{k}^{T} &= \dfrac{\left(\rho_{u} \left( \sum\limits_{m=1}^{M} \left(\sum\limits_{n=1}^{N} x_{mn}  \right) \gamma_{mk}\right)^{2} -  T    \sum\limits_{m=1}^{M} \left(\rho_{u}\sum\limits_{i=1}^{K}\left(\beta_{mi} - \gamma_{mi} \right) + 1 \right) \left(\sum\limits_{n=1}^{N} x_{mn}  \right) \gamma_{mk}\right)}{T \rho_{u}}.
			\end{aligned}
		\end{equation}
	\end{lemma}
	\begin{proof}
		Please refer to Appendix \ref{App: CFOP_ProofUnivariateLemma} for the proof.
	\end{proof}
	To evaluate (\ref{Eq: CFOP_PoutOrthoSim2}), we need to solve a $MN$th order integration. For typical values of $M$ and $N$ used in cell-free massive MIMO systems, say $M = 32$ and $N=1$, it is intractable to solve a $32$th order integration even in popular software such as \textit{Matlab}, \textit{Mathematica}, etc. Thus, it is important to approximate (\ref{Eq: CFOP_PoutOrthoSim2}) for evaluation and analysis. To circumvent the intractability, we propose to utilize the uni-variate approximation from \cite{Rahman2004:Integral_DimensionReduction}. Using this method, one can tightly approximate an $MN$th order integration by a sum of $MN$ single-order integration. The approximation and detailed proof are presented in the following theorem.
	\begin{theorem}\label{Thm: CFOP_OPUnivariate}
		For the case of no pilot contamination, the OP of $ k $th user for threshold $ T $ is approximated as
		\begin{footnotesize}
			\begin{equation}\label{Eq: CFOP_Pout_NoPilot_AppFinal}
				\begin{aligned}
					P_{out}^{k}\left(T\right) &\approx   1 - N\sum_{m=1}^{M}\int_{0}^{\infty}\left( \sum\limits_{\substack{i=1 \\ i\ne k}}^{K} \left(\frac{\left(x C_{1,m}^{i} + C_{2,m}^{i}\right)^{K-2}}{\prod\limits^{K}_{{\substack{j=1 \\ j \ne i,j \ne k}}} x C_{3,m}^{i,j} + C_{4,m}^{i,j}} \left[ 1 - e^{-\left(\frac{x^{2} C_{5,m} + x C_{9,m} + C_{10,m}}{ x C_{1,m}^{i} + C_{2,m}^{i}}\right)}  \right]\right)\operatorname{U}\left(x^{2} C_{5,m} + xC_{9,m} + C_{10,m}\right) \right) e^{-x} dx  
					\\ &+ (MN - 1)\left( \sum\limits_{\substack{i=1 \\ i\ne k}}^{K}  \frac{C_{i}^{K-2}}{\prod\limits^{K}_{{\substack{j=1 \\ j \ne i,j \ne k}}} \left(C_{i} - C_{j} \right)} \left[ 1 -e^{-\frac{C_{k}^{T}}{C_{i}}}  \right] \right)\operatorname{U}\left(C_{k}^{T}\right).
				\end{aligned}
			\end{equation}
		\end{footnotesize}
	\end{theorem}
	\begin{proof}
		The derivation details are given in Appendix \ref{App: CFOP_ProofUnivariateTheorem}.
	\end{proof}
	\subsection{Single-cell collocated mMIMO}
	Single-cell collocated mMIMO can be considered as the special case of CF-mMIMO system, when all the $M$ APs are collocated also termed as base station (BS), and $ K $ pilot sequences are pairwise orthogonal then we have  $\beta_{mk} = \beta_{m^{\prime}k} \triangleq \beta_{k}$, $ \gamma_{mk} = \gamma_{m^{\prime} k} \triangleq \gamma_{k}$ and no pilot contamination. The Corollary \ref{Cor: CFOP_LogNormalNPC} and Theorem \ref{Thm: CFOP_OPUnivariate} are applicable to calculate the OP for this special case. However, the integral involved in \eqref{Eq: CFOP_Pout_NoPilot_AppFinal} is simple enough to be solved in closed form. For a fair comparison, we consider $MN$ antenna single-cell collocated mMIMO system so that the total number of antenna remains same in both CF-mMIMO and collocated mMIMO system. The following corollaries present the OP and rate approximations for the single-cell collocated mMIMO system. 
	
	\begin{corollary}[Corollary to Theorem \ref{Thm: CFOP_SINRLogNormal}]\label{Cor: CFOP_Pout_LogNormal_mMIMO}
		For the single cell collocated mMIMO scenario, the OP of $ k $th user is approximated using \eqref{Eq: CFOP_POutLogNormalPC}, where \eqref{Eq: CFOP_MuLogNormalPC} and \eqref{Eq: CFOP_SigmaLogNormalPC} are evaluated using the following simplified expression for the moments.
		\begin{equation}\label{Eq: CFOP_XMean_mMIMO}
			\begin{aligned}
				\mathbb{E}\left[X_{u,k}\right] 
				&= \rho_{u}\left( MN\right)_{2}\gamma_{k}^{2}.
			\end{aligned}
		\end{equation}
		\begin{equation}\label{Eq: CFOP_X2Mean_mMIMO}
			\begin{aligned}
				\mathbb{E}\left[X_{u,k}^{2} \right] &= \rho_{u}^{2} \left( M N \right)_{4}\gamma_{k}^{4}. 
			\end{aligned}
		\end{equation}
		\begin{equation}\label{Eq: CFOP_YMean_mMIMO_NPC}
			\begin{aligned}
				\mathbb{E}\left[Y_{u,k}\right] &= MN\gamma_{k} \left[ \rho_{u}  \left(\sum\limits_{i\ne k}^{K} \gamma_{i} \right) + 1 +  \rho_{u} \sum_{i=1}^{K}\left(\beta_{i} - \gamma_{i}\right) \right].
			\end{aligned}
		\end{equation}
		\begin{equation}\label{Eq: CFOP_Y2Mean_mMIMO_NPC}
			\begin{aligned}
				\mathbb{E}\left[Y_{u,k}^{2}\right] &= \left( M N \right)_{2} \gamma_{k}^{2} \left[  \rho_{u}^{2}\left( \sum_{i\ne k}^{K}\gamma_{i}^{2} + \left(\sum_{i\ne k}^{K}\gamma_{i}\right)^{2} \right)  + 1 + \rho_{u}^{2} \sum_{i=1}^{K}\left(\beta_{i} - \gamma_{i}\right)^{2} \right. \\
				& \left.  + 2\rho_{u} \left(\sum_{i\ne k}^{K}\gamma_{i} 
				\right) + 2\rho_{u}  \sum_{i=1}^{K}\left(\beta_{i} - \gamma_{i}\right) + 2\rho_{u}^{2} \sum_{i=1}^{K}\left(\beta_{i} - \gamma_{i}\right) \left(\sum_{i\ne k}^{K} \gamma_{i} \right) \right].
			\end{aligned}
		\end{equation}
		\begin{equation}\label{Eq: CFOP_CorrXY_mMIMO_NPC}
			\begin{aligned}
				\mathbb{E}\left[X_{u,k}Y_{u,k}\right] 
				&= \rho_{u}\left(M N\right)_{3} \gamma_{k}^{3} \left[ \rho_{u} \left(\sum_{i\ne k}^{K}  \gamma_{i}\right) + 1 + \rho_{u}\sum_{i=1}^{K}\left(\beta_{i} - \gamma_{i}\right)\right].
			\end{aligned} 
		\end{equation}
	\end{corollary}
	\begin{corollary}[Corollary to Theorem \ref{Thm: CFOP_RateLogNormal}]\label{Cor: CFOP_Rate_LogNormal_mMIMO}
		For the single cell collocated mMIMO scenario, the ergodic rate and the respective lower and upper bounds of $ k $-th user are computed using the \eqref{Eq: CFOP_ER_PC} and  \eqref{Eq: CFOP_ERBs_PC} with the parameters of Log-normal are calculated using \eqref{Eq: CFOP_XMean_mMIMO} - \eqref{Eq: CFOP_CorrXY_mMIMO_NPC}.
	\end{corollary}
	\begin{corollary}[Corollary to Theorem \ref{Thm: CFOP_OPUnivariate}]\label{Cor: CFOP_Pout_NPC_mMIMO}
		For the single cell collocated mMIMO scenario, the OP of $k$th user is approximated as  
		\begin{equation}\label{Eq: CFOP_Pout_NPCmMIMO_Case1}
			\begin{aligned}
				P_{out}^{k}\left(T\right) &\approx 1 - M N\sum\limits_{\substack{i=1 \\ i\ne k}}^{K} D_{1}^{i}\left[1 -  \frac{e^{-D_{3}^{i}}}{D_{2}^{i} + 1}\right] + \left(MN -1 \right)\sum\limits_{\substack{i=1 \\ i\ne k}}^{K} D_{1}^{i}\left[ 1 -e^{- \left( D_{2}^{i} + D_{3}^{i}\right)}  \right]
			\end{aligned}
		\end{equation}
		for $ T \le \frac{\rho_{u} \left( MN-1 \right)\gamma_{k}}{\left( \rho_{u} \sum\limits_{i=1}^{K}\left(\beta_{i} - \gamma_{i} \right) + 1\right)} $, and 
		\begin{equation}\label{Eq: CFOP_Pout_NPCmMIMO_Case2}
			\begin{aligned}
				P_{out}^K(T) &\approx 1 - MN \sum\limits_{\substack{i=1 \\ i\ne k}}^{K} D_{1}^{i}\left[ e^{-\kappa}  - \frac{e^{-D_{3}^{i}} e^{-\kappa (D_{2}^{i} + 1)}}{D_{2}^{i} + 1}\right] \\ &+ \left(MN-1\right)\sum\limits_{\substack{i=1 \\ i\ne k}}^{K} D_{1}^{i}\left[ 1 -e^{- \left( D_{2}^{i} + D_{3}^{i}\right)}  \right]\operatorname{U}\left(D_{4} + D_{5} + D_{6}\right)
			\end{aligned}
		\end{equation}
		for $ T > \frac{\rho_{u} \left( MN-1 \right)\gamma_{k}}{\left( \rho_{u} \sum\limits_{i=1}^{K}\left(\beta_{i} - \gamma_{i} \right) + 1\right)} $, where $D_{1}^{i},D_{2}^{i},D_{3}^{i},D_{4},D_{5}$, $D_{6}$ and $\kappa $ are defined in (\ref{Eq: D1D2D3mMIMO}), (\ref{Eq: D4D5D6mMIMO}) and \eqref{Eq: kappamMIMO}.
	\end{corollary}
	\begin{proof}
		Using the fact that $\beta_{mk} = \beta_{m^{\prime}k} \triangleq \beta_{k}, \gamma_{mk} = \gamma_{m^{\prime} k} \triangleq \gamma_{k}$. We have simplified the \eqref{Eq: CFOP_Pout_NoPilot_AppFinal}. The details are provided in Appendix \ref{App: CFOP_UnivariatemMIMOCor}
	\end{proof}
	Note that the expressions in Corollary \ref{Cor: CFOP_Pout_NPC_mMIMO} are in closed form and do not require any numerical integration.  The obtained expressions for the single-cell collocated mMIMO system with imperfect CSI are novel to the best of our knowledge.
	
	\section{Results \& Discussion}\label{Sec: CFOP_Outage_Results}
	The simulation setup is similar to that in \cite{Ngo2017:CellFree} and is repeated here for completeness. A cell-free massive MIMO system with various $M$ and $K$ values have been considered. All $M$ AP and $K$ users are dispersed in a square of area $D \times D \: \text{km}^{2}$.  The large-scale fading coefficient $\beta_{mk}$ models the path loss
	and shadow fading according to
	\begin{equation}
		\beta_{mk}= PL_{mk}10^{\frac{\sigma_{th}z_{mk}}{10}},
	\end{equation}
	where $PL_{mk}$ represents the path loss, $\sigma_{th}$ represents the standard deviation of the shadowing and $z_{mk} \sim \mathcal{N }(0, 1)$. The relation between the path loss $PL_{mk}$ and the distance between the distance $d_{mk}$ between the $m$th AP and $k$th user is obtained using the three slope model in \cite[Eq. 52]{Ngo2017:CellFree}. The other parameters are summarized in Table \ref{Tab:CFOP_params}.  The normalized transmit SNRs $ \rho_{p}$ and $ \rho_{u}$ are obtained by dividing the actual transmit powers $\bar \rho_{p}$ and $\bar \rho_{u}$ by the noise power, respectively.
	\begin{table}[!t]
		\centering
		\begin{tabular}{|l|r|}
			\hline
			Parameter & value \\
			\hline
			\hline
			Carrier frequency & $1.9~GHz$ \\
			\hline
			Bandwidth & $20~MHz$\\
			\hline
			Noise figure & $9$ dB\\
			\hline
			AP antenna height & $15~m$\\
			\hline
			User antenna height & $1.65~m$\\
			\hline
			$\sigma_{sh}$ & $8$ dB\\
			\hline
			$\bar \rho_p$, $\bar \rho_u$ & $100~mW$\\
			\hline
		\end{tabular}
		\caption{Simulation parameters}
		\label{Tab:CFOP_params}
	\end{table}
	\begin{figure}[!t]
		\begin{subfigure}{0.48\textwidth}
			\includegraphics[width=\textwidth]{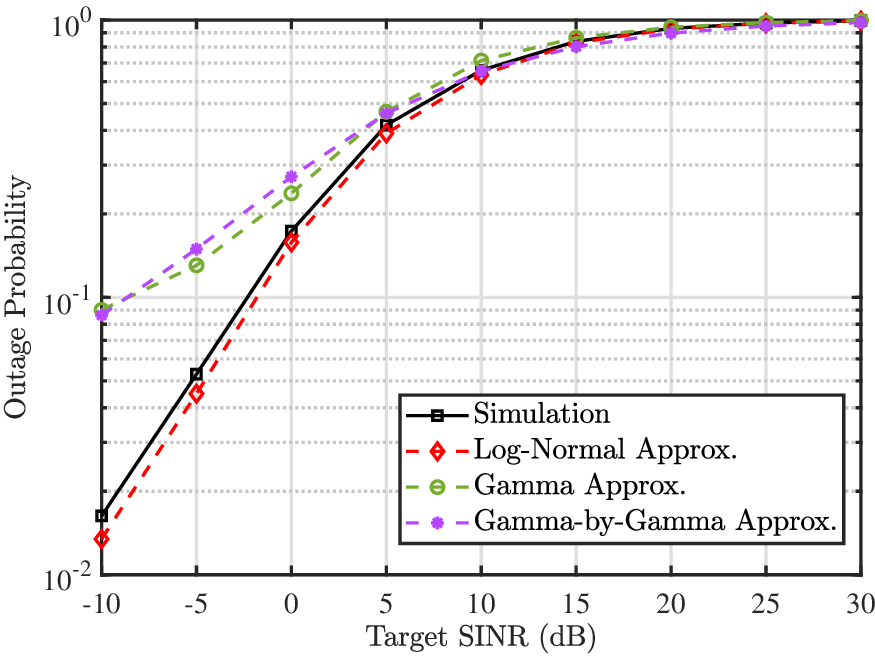}
			\caption{With Pilot Contamination ($ M = 300, N = 1 $ and $ K = 30 $) }
		\end{subfigure}
		\hspace{3mm}
		\begin{subfigure}{0.48\textwidth}
			\includegraphics[width=\textwidth]{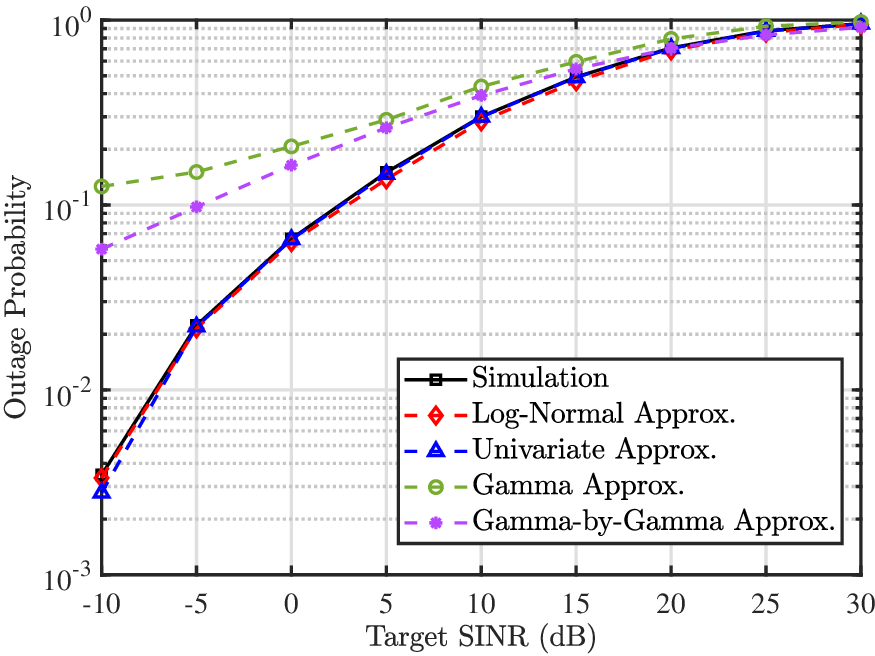}
			\caption{Without pilot contamination ($ M = 100, N = 1 $ and $ K = 5 $)}
		\end{subfigure}
		\caption{Comparison of OP using different existing approaches with the proposed methods}
		\label{Fig:CFOPComp_DiffApproaches}
	\end{figure} 
	\par We first compare our OP approximation in Theorem \ref{Thm: CFOP_SINRLogNormal} and \ref{Thm: CFOP_OPUnivariate} with the existing moment-matching approximation approaches in Fig. \ref{Fig:CFOPComp_DiffApproaches}. Two different moment-matching approaches are compared with our approximation. In the first, the SINR is approximated by a Gamma RV following \cite{srinivasan2019analysis}, whereas, in the second, the numerators and denominators of the SINR are separately approximated by Gamma RVs. Then the ratio of two Gamma RV is Beta-prime RV \cite{cordeiro2012mcdonald}. Note that the Gamma approximation and beta-prime approximation fail to capture the tail behavior of OP in both scenarios \textit{i.e.,} with and without pilot contamination. This is because the single Gamma approximation relies on the approximate first and the second moment of SINR obtained using bi-variate Taylor series expansion, which does not provide a good approximation of SINR's moments for the CF-mMIMO system. Next, the Gamma-by-Gamma or the Beta-prime approximation fails due to the correlation between the numerator and denominator of SINR owing to the use of MRC. Other treatises such as \cite{Ding2018:OutageADCMassiveMIMO} approximate the OP by proving that SCV of all but one component of the SINR is zero and obtaining OP by transforming the CDF of the remaining term. For our case, through extensive simulations, we determined that the SCV of more than one component of SINR is non-zero, and hence the method cannot be applied.  These results justify the necessity of the new approximations proposed in this work.  In the subsequent subsections, we investigate the OP performance of CF-mMIMO (with and without pilot contamination) for various values of $M$ and $K$. The numerical results are generated as follows.
	\begin{enumerate}
		\item We realized $100$ random deployment of APs and UEs. The large-scale fading coefficients, with the shadowing effect, are calculated for each realization.
		\item For each realization, $10,000$ Monte Carlo iterations are performed to calculate the SINR for each UE. Using this SINR, we calculated the OP and ergodic rate of the UEs.
		\item System's OP and the system's ergodic rate are the averages of the OP and ergodic rate in each realization.
	\end{enumerate}
	
	\subsection{Results for With Pilot Contamination}
\begin{figure}[!t]
		\begin{subfigure}{0.48\textwidth}
			\includegraphics[width=\textwidth]{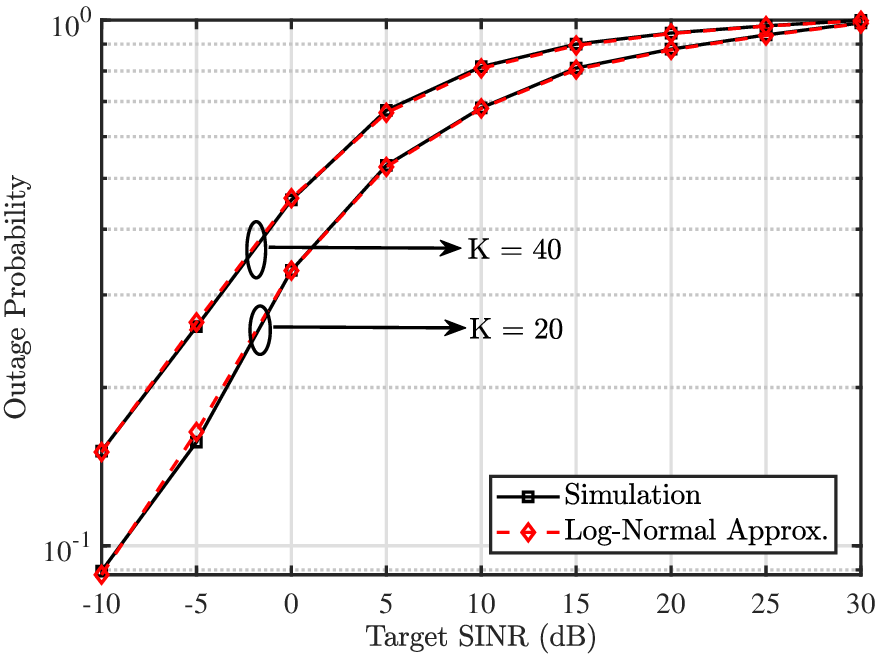}
			\caption{For fixed $M = 80, N = 4$.}
			\label{Fig: CFOP_PC_OP_Fixed_M}
		\end{subfigure}
		\hspace{3mm}
		\begin{subfigure}{0.48\textwidth}
			\includegraphics[width=\textwidth]{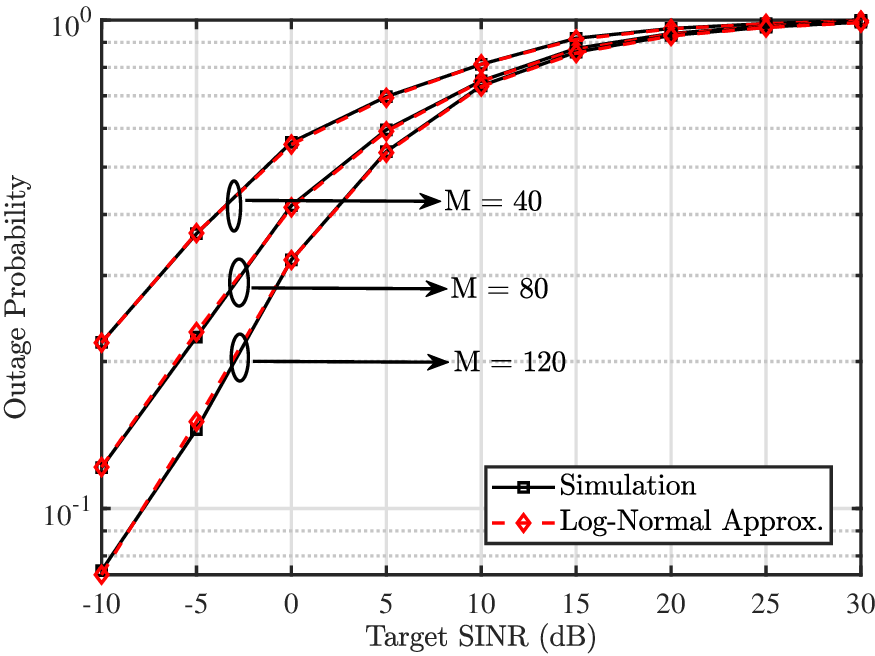}
			\caption{For fixed $K = 30, N = 4$.}
			\label{Fig: CFOP_PC_OP_Fixed_K}
		\end{subfigure}
		\caption{Impact of $M$ and $K$ on the OP of CF-mMIMO system with pilot contamination}
		\label{Fig:CFOP_PC_OP_Results_VaryM_K}
	\end{figure}

\begin{figure}[!t]
		\begin{subfigure}{0.48\textwidth}
			\includegraphics[width=\textwidth]{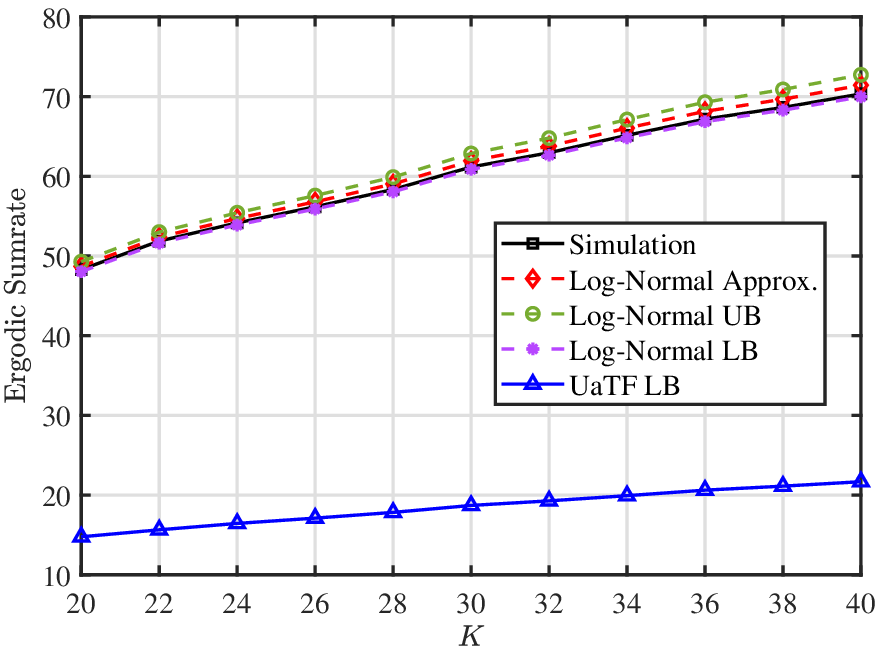}
			\caption{For fixed $M = 80, N = 4$.}
			\label{Fig:CFOP_PC_Rate_Fixed_M}
		\end{subfigure}
		\hspace{3mm}
		\begin{subfigure}{0.48\textwidth}
			\includegraphics[width=\textwidth]{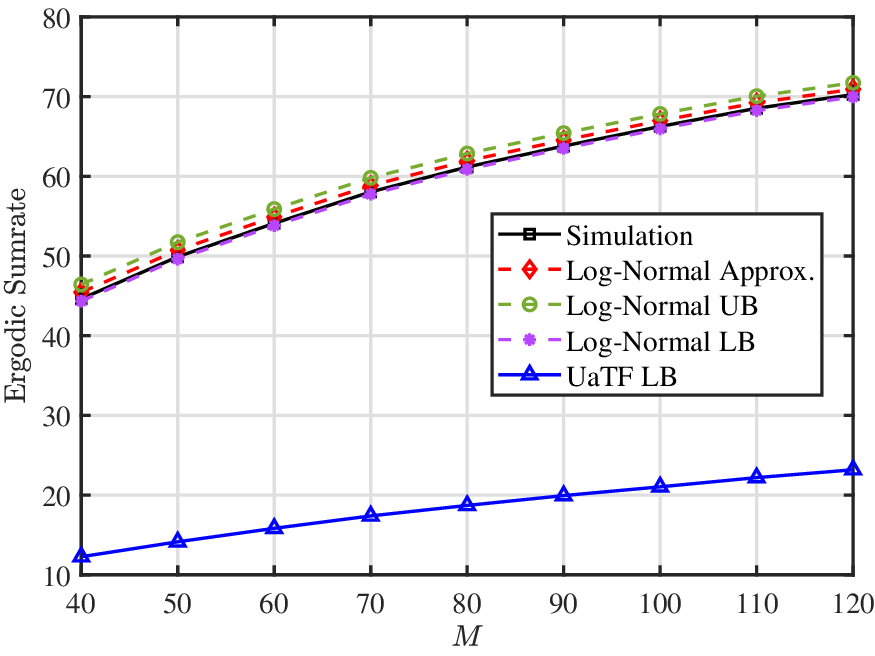}
			\caption{For fixed $K = 30, N = 4$.}
			\label{Fig:CFOP_PC_Rate_Fixed_K}
		\end{subfigure}
		\caption{Impact of $M$ and $K$ on the ergodic rate of CF-mMIMO system with pilot contamination}
		\label{Fig:CFOP_PC_Rate_Results_VaryM_K}
	\end{figure}

	In this sub-section, the expression in Theorem \ref{Thm: CFOP_SINRLogNormal} and \ref{Thm: CFOP_RateLogNormal} are validated through numerical simulation. Fig. \ref{Fig: CFOP_PC_OP_Fixed_M} and \ref{Fig: CFOP_PC_OP_Fixed_K} show the trend of OP for varying $K$ and $M$, respectively. The number of antennas per AP \textit{i.e.,} $N$ is chosen to be $4$. It is evident that the proposed two-step Log-normal approximation is closely matching with the simulated OP. It is clear from Fig \ref{Fig: CFOP_PC_OP_Fixed_M} that the OP increases as the number of users increase due to the corresponding increase in pilot contamination and interference. Similarly, the OP decreases with the increase in the number of APs in the system, as shown in Fig \ref{Fig: CFOP_PC_OP_Fixed_K} for fixed $ K = 30 $ and $N=4$.    
	Fig. \ref{Fig:CFOP_PC_Rate_Fixed_M} and \ref{Fig:CFOP_PC_Rate_Fixed_K} shows the simulated values for ergodic sumrate and the approximate sumrate obtained using Theorem \ref{Thm: CFOP_RateLogNormal} for CF-mMIMO with pilot contamination. It can be observed that the popular use-and-then-forget (UaTF) lower bound (LB) severely underestimates the ergodic sumrate as compared to the simulated one. Theorem \ref{Thm: CFOP_RateLogNormal} proposed an ergodic rate expression in terms of an integral and also provided simple and closed-form lower and upper bound for the proposed integral. The approximate rate values and bounds better match the simulated one compared to UaTF LB.   
	
	\subsection{Results for Without Pilot Contamination}
\begin{figure}[!t]
		\begin{subfigure}{0.48\textwidth}
			\includegraphics[width=\textwidth]{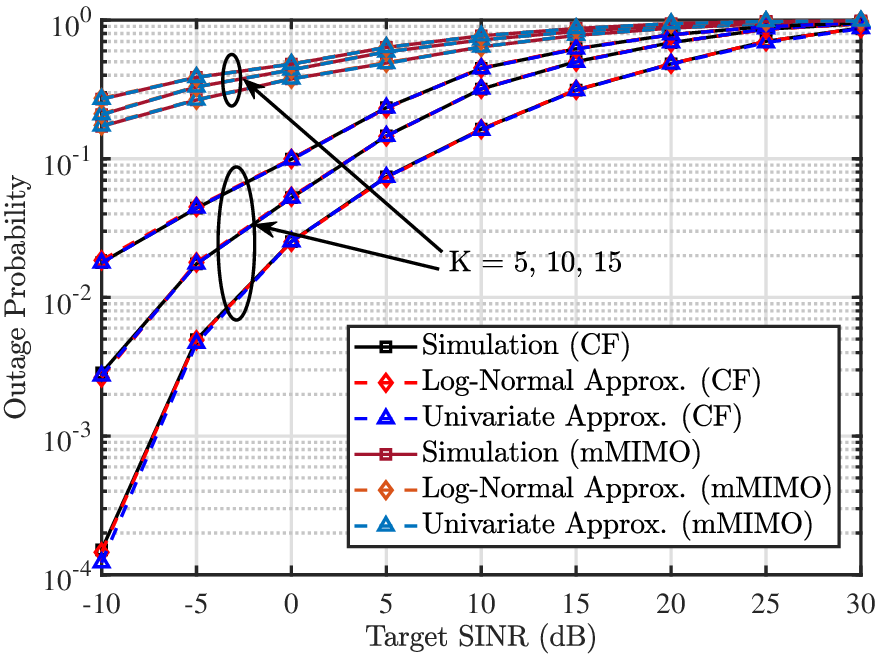}
			\caption{For fixed $M = 80, N = 4$.}
			\label{Fig: CFOP_NPC_OP_Fixed_M}
		\end{subfigure}
		\hspace{3mm}
		\begin{subfigure}{0.48\textwidth}
			\includegraphics[width=\textwidth]{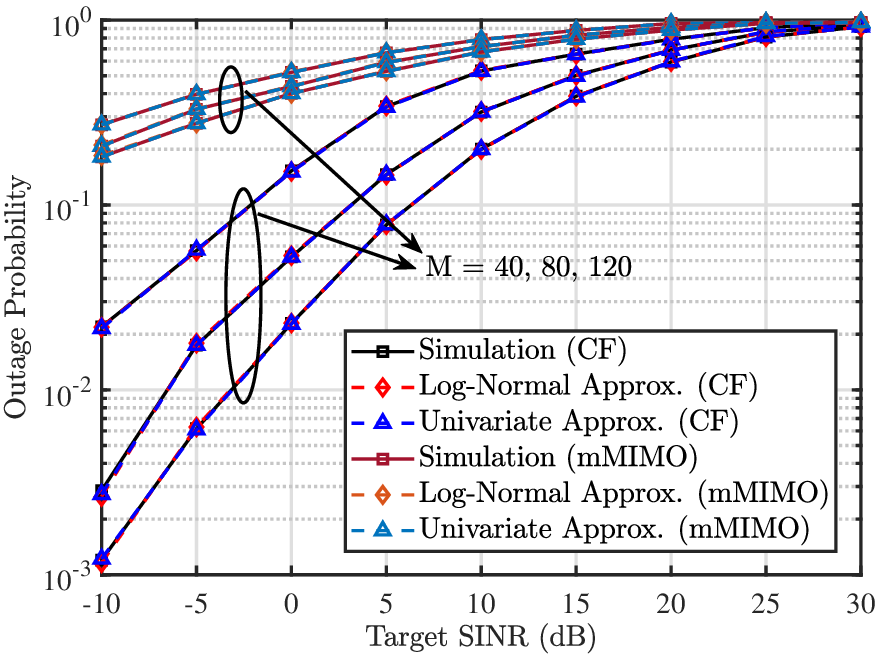}
			\caption{For fixed $K = 10, N = 4$.}
			\label{Fig: CFOP_NPC_OP_Fixed_K}
		\end{subfigure}
		\caption{Impact of $M$ and $K$ on the OP of CF-mMIMO and mMIMO system without pilot contamination}
		\label{Fig:CFOP_NPC_OP_Results_VaryM_K}
	\end{figure}

\begin{figure}[!t]
		\begin{subfigure}{0.48\textwidth}
			\includegraphics[width=\textwidth]{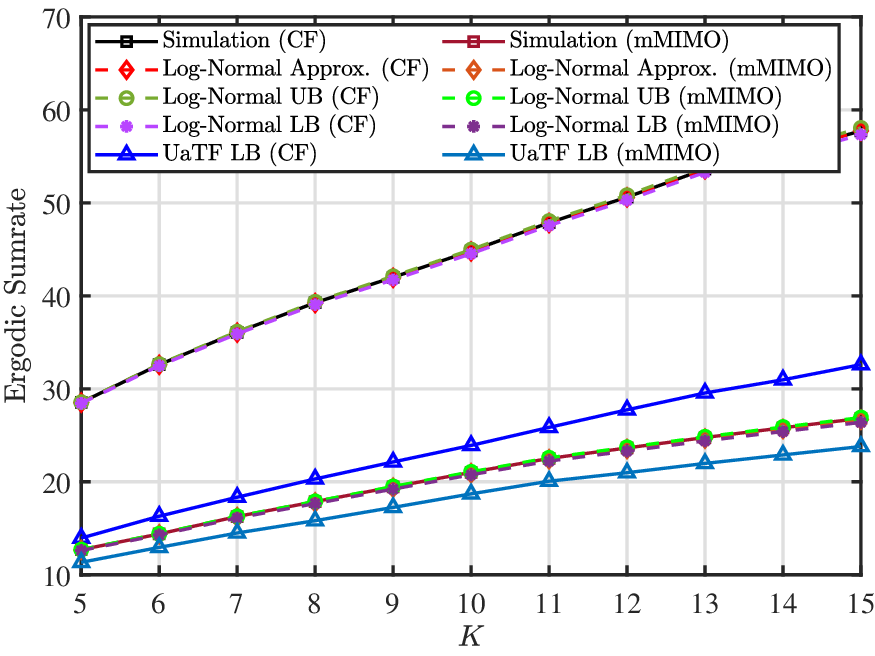}
			\caption{For fixed $M = 80, N = 4$.}
			\label{Fig: CFOP_NPC_Rate_Fixed_M}
		\end{subfigure}
		\hspace{3mm}
		\begin{subfigure}{0.48\textwidth}
			\includegraphics[width=\textwidth]{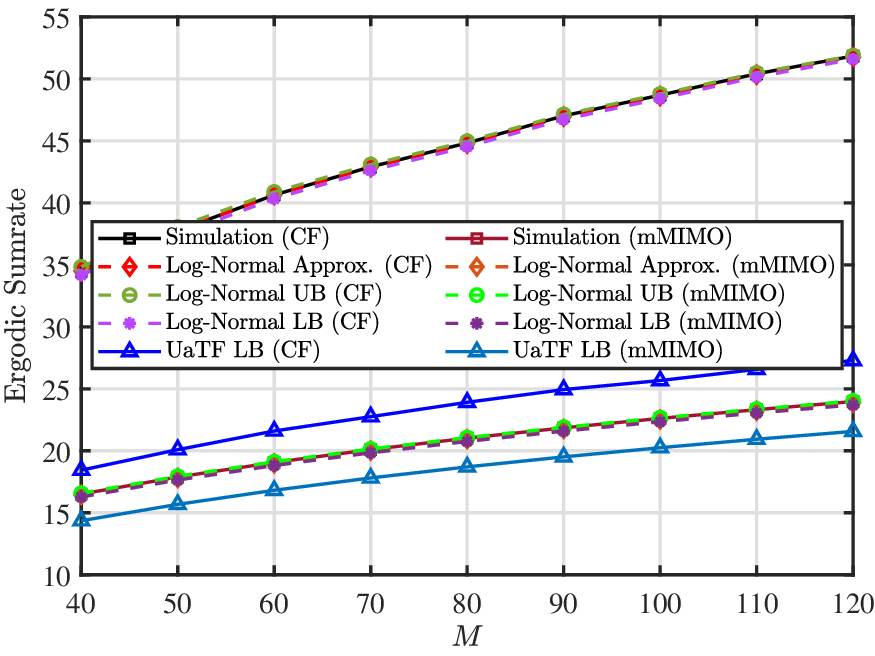}
			\caption{For fixed $K = 10, N = 4$.}
			\label{Fig: CFOP_NPC_Rate_Fixed_K}
		\end{subfigure}
		\caption{Impact of $M$ and $K$ on the ergodic rate of CF-mMIMO and mMIMO system without pilot contamination}
		\label{Fig:CFOP_NPC_Rate_Results_VaryM_K}
	\end{figure}

	In this sub-section, we compare the performance of CF-mMIMO and the single-cell collocated  mMIMO systems without pilot contamination. Fig. \ref{Fig: CFOP_NPC_OP_Fixed_M} and \ref{Fig: CFOP_NPC_OP_Fixed_K} shows that the approximation presented in Theorem \ref{Thm: CFOP_OPUnivariate}, Corollary \ref{Cor: CFOP_LogNormalNPC} and Corollary \ref{Cor: CFOP_Pout_LogNormal_mMIMO}, \ref{Cor: CFOP_Pout_NPC_mMIMO} for CF-mMIMO and mMIMO, respectively are closely matching with simulation results. Here, also we observed that not only does the CF-mMIMO performs better than mMIMO, but the improvement it shows with varying parameter is also more significant than the mMIMO. For example, when $M$ increases to $80$ from $40$ at target SINR of $-5$ dB and $N=4$, OP decreases by $69.45 \%$ for the CF-mMIMO system, whereas it decreases by only $16.61 \% $ for the mMIMO system when antennas are increased from $160$ to $320$. Hence, it is better to increase the density of APs as compared to increasing the antenna at a single collocated AP. 
	Next, Fig. \ref{Fig: CFOP_NPC_Rate_Fixed_M} and \ref{Fig: CFOP_NPC_Rate_Fixed_K} present the ergodic sumrate of CF-mMIMO and mMIMO system without pilot contamination. Again, it is observed that the ergodic sumrate calculated using Corollary \ref{Cor: CFOP_NPC_Rate_LogNormal} provides a better estimate for sumrate as compared to the UaTF bound, which heavily underestimates the performance of CF-mMIMO as well as mMIMO system. Also, the bounds provided for the integral in \eqref{Eq: CFOP_ER_PC} tightly bounds it and are easy to compute as expressions are available in closed form. 
	\subsection{Correlated Rician fading scenario} For the case of correlated Rician fading the channel between $m$th AP and $k$th is modeled as
	\begin{equation}
		\begin{aligned}
			\mathbf{g}_{mk} \sim \mathcal{CN}\left(\bar{\mathbf{g}}_{mk},\mathbf{R}_{mk}\right),
		\end{aligned}    
	\end{equation}
	where $\bar{\mathbf{g}}_{mk} \in \mathbb{C} $ is the LOS component and, $\mathbf{R}_{mk} \in \mathbb{C}^{N\times N}$ is the spatial correlation matrix between $m$th AP and $k$th user. Following from \cite{Bjornson2020:CompetitiveCellFree,Ngo2018:Rician_CellFree,Emil2019:RicianTWC}, the LMMSE estimate of channel vector is $ \hat{\mathbf{g}}_{mk} \sim \mathcal{CN}\left( \bar{\mathbf{g}}_{mk}, \mathbf{R}_{mk} \mathbf{C}^{-1}_{mk} \mathbf{R}_{mk} \right), $ where $\mathbf{C}_{mk} = \tau_{p}\rho_{p}\sum\limits_{i=1}^{K} \mathbf{R}_{mi} \abs{\boldsymbol{\phi}_{k}^{H} \boldsymbol{\phi}_{i}}^{2} + \mathbf{I}_{N}$ and the channel estimation error \textit{i.e.}, $ \boldsymbol{\varepsilon}_{mk} = \mathbf{g}_{mk} - \hat{\mathbf{g}}_{mk} \sim \mathcal{CN}\left(\mathbf{0},\boldsymbol{\Lambda}_{mk} \right)$. The effective  SINR of the $ k $-th user is as follows:
	\begin{equation}\label{Eq: CFOP_SINRInnerCorr}
		\begin{aligned}
			\lambda_{u,k} = \frac{X_{u,k}}{Y_{u,k}} 
			= \frac{\rho_{u} \norm*{\hat{\mathbf{g}}_{k}}^{4} }{ \rho_{u} \sum\limits_{i\ne k}^{K} \abs*{\mathbf{\hat{g}}_{k}^{H} \mathbf{\hat{g}}_{i}}^{2}  +   \hat{\mathbf{g}}_{k}^{H}\left( \rho_{u}\sum\limits_{i=1}^{K} \boldsymbol{\Lambda}_{i} + \mathbf{I}_{MN} \right)\hat{\mathbf{g}}_{k} },
		\end{aligned} 
	\end{equation}
	where $  \hat{\mathbf{g}}_{k} = \begin{bmatrix}
		\hat{\mathbf{g}}_{1k} \ \dots \ \hat{\mathbf{g}}_{Mk} 
	\end{bmatrix}^{T} $ and $ \boldsymbol{\Lambda}_{i} = \operatorname{diag}\left(\boldsymbol{\Lambda}_{1i},\dots,\boldsymbol{\Lambda}_{Mi} \right) $. Note that the SINR expression is functionally the same as for the independent Rayleigh fading case. Hence, the two-step moment matching method can also be extended for the case of correlated Rician fading. The major difference is that the calculation of moments of the numerator and denominator will be more involved algebraically and is beyond the scope of the paper.
	\section{Conclusion}\label{Sec: CFOP_Conclusion}
	In this work, we derived approximate OP and rate expressions for a CF-mMIMO system with and without pilot contamination under the Rayleigh faded channels. We used the two-step moment matching to derive the approximate expressions and provided simple expressions for the OP and ergodic rate. In the case of no pilot contamination, an exact expression is derived in terms of a multi-fold integral. A simple and accurate approximation using the uni-variate dimension reduction method is proposed to circumvent the evaluation of higher-order integration. Specific to the single-cell collocated mMIMO system, approximate OP expressions are obtained in closed form, which involves only elementary functions. The validity of the approximations, derived for both CF-mMIMO and mMIMO, was verified by Monte-Carlo simulations. Investigating the effect of correlated fading with a line-of-sight component will be an interesting future direction.
	
	\appendices
	\section{Proof for Theorem \ref{Thm: CFOP_SINRLogNormal}}\label{App: CFOP_ProofSINRLogNormal}
	Using technique of moment matching, we first approximate the $ X_{u,k} $ and $ Y_{u,k} $ by Log-normal distribution \textit{i.e.,} $ X_{u,k} \sim \operatorname{LN}\left(\mu_{X_{u,k}},\sigma_{X_{u,k}}^{2}\right) $ and $ Y_{u,k} \sim \operatorname{LN}\left(\mu_{Y_{u,k}},\sigma_{Y_{u,k}}^{2}\right) $ with parameter given as
	\begin{equation}\label{Eq:CFOP_XMuSigma}
		\begin{aligned}
			\mu_{X_{u,k}} &= \log\left(\frac{\left(\mathbb{E}\left[X_{u,k} \right]\right)^{2}}{\sqrt{\mathbb{E}\left[ X_{u,k}^{2}\right]}}\right)  \quad \sigma_{X_{u,k}} &= \sqrt{\log\left( \frac{\mathbb{E}\left[ X_{u,k}^{2}\right]}{\left(\mathbb{E}\left[X_{u,k} \right]\right)^{2}}\right)},
		\end{aligned}
	\end{equation}
	and 
	\begin{equation}\label{Eq:CFOP_YMuSigma}
		\begin{aligned}
			\mu_{Y_{u,k}} &= \log\left(\frac{\left(\mathbb{E}\left[Y_{u,k} \right]\right)^{2}}{\sqrt{\mathbb{E}\left[ Y_{u,k}^{2}\right]}}\right) \quad \sigma_{Y_{u,k}} &= \sqrt{\log\left( \frac{\mathbb{E}\left[ Y_{u,k}^{2}\right]}{\left(\mathbb{E}\left[Y_{u,k} \right]\right)^{2}}\right)}.
		\end{aligned}
	\end{equation}	 
	The parameters in (\ref{Eq:CFOP_XMuSigma}) and (\ref{Eq:CFOP_YMuSigma}) can be evaluated using (\ref{Eq:CFOP_XMean}), (\ref{Eq:CFOP_X2Mean}), (\ref{Eq:CFOP_YMean}) and (\ref{Eq:CFOP_Y2}).  Consider
	\begin{equation}
		\begin{aligned}
			\log\left(\lambda_{u,k}\right)	&= \log\left(X_{u,k}\right) - \log\left(Y_{u,k}\right).
		\end{aligned}
	\end{equation}
	Under Log-normal assumption, $ \log\left(X_{u,k}\right) $ and $ \log\left(Y_{u,k}\right) $ follow normal distribution, \textit{i.e.}, $ \log\left(X_{u,k}\right) \sim \mathcal{N}\left(\mu_{X_{u,k}},	\sigma^{2}_{X_{u,k}}\right) $ and $ \log\left(Y_{u,k}\right) \sim \mathcal{N}\left(\mu_{Y_{u,k}},	\sigma^{2}_{Y_{u,k}}\right) $, hence, $ \log\left(\lambda_{u,k}\right) $ follows normal distribution with 
	\begin{equation}
		\begin{aligned}
			\mu_{\lambda_{u,k}}	&= \mathbb{E}\left[\log\left(X_{u,k}\right)\right] - \mathbb{E}\left[\log\left(Y_{u,k}\right)\right], \\
			&= \mu_{X_{u,k}} - \mu_{Y_{u,k}}.
		\end{aligned}
	\end{equation}
	and
	\begin{equation}
		\begin{aligned}
			\sigma_{\lambda_{u,k}}^{2}	&= \mathbb{V}\left[\log\left(X_{u,k}\right)\right] +  \mathbb{V}\left[\log\left(Y_{u,k}\right)\right] - 2 \operatorname{Cov}\left(\log\left(X_{u,k}\right),\log\left(Y_{u,k}\right)\right),
		\end{aligned}
	\end{equation}
	where $ \operatorname{Cov}\left(\log\left(X_{u,k}\right),\log\left(Y_{u,k}\right)\right) $ is \cite{vzerovnik2013transformation},
	\begin{equation}
		\begin{aligned}		         \operatorname{Cov}\left(\log\left(X_{u,k}\right),\log\left(Y_{u,k}\right)\right) &= \log\left( \frac{\operatorname{Cov}\left(X_{u,k},Y_{u,k}\right)}{\mathbb{E}\left[X_{u,k}\right]\mathbb{E}\left[Y_{u,k}\right]} + 1\right). 
		\end{aligned}
	\end{equation}
	So, 	\begin{equation}
		\begin{aligned}
			\sigma_{\lambda_{u,k}}^{2}	&= \sigma^{2}_{X_{u,k}} +  \sigma^{2}_{Y_{u,k}} - 2 \log\left( \frac{\mathbb{E}\left[X_{u,k}Y_{u,k}\right]}{\mathbb{E}\left[X_{u,k}\right]\mathbb{E}\left[Y_{u,k}\right]}\right),
		\end{aligned}
	\end{equation}
	where the $ \mathbb{E}\left[X_{u,k}Y_{u,k}\right] $ is calculated using (\ref{Eq:CFOP_CorrXY}). This completes the proof.
	\section{Moments Calculation}
	Here, we have derived the first and second moment of the RV $ X_{u,k} $ and $ Y_{u,k} $. From (\ref{Eq: CFOP_SINRInner}), we have 
	\begin{equation}\label{Eq:CFOP_XMean}
		\begin{aligned}
			\mathbb{E}\left[X_{u,k}\right] &= \rho_{u}\mathbb{E}\left[ \left( \sum\limits_{m=1}^{M}  \norm*{\hat{\mathbf{g}}_{mk}}^{2} \right)^{2}\right] = \rho_{u}\mathbb{E}\left[\sum_{m = 1}^{M}\norm*{ \hat{g}_{mk}}^{4} + \sum_{m = 1}^{M}\sum_{n \ne m}^{M} \norm*{ \hat{g}_{mk}}^{2} \norm*{\hat{g}_{nk}}^{2} \right], \\
			&= \rho_{u}N\left[\sum_{m = 1}^{M} \gamma_{mk}^{2} + N\left(\sum_{m = 1}^{M} \gamma_{mk}\right)^{2}\right].
		\end{aligned}
	\end{equation}
	Similarly, after squaring the $ X_{u,k} $ and taking term-by-term expectations, we have  
	\begin{equation}\label{Eq:CFOP_X2Mean}
		\begin{aligned}
			\mathbb{E}\left[X_{u,k}^{2}\right] &= \left(\mathbb{E}\left[X_{u,k}\right]\right)^{2} + \rho_{u}^{2}\left( 6N\sum_{m=1}^{M}  \gamma_{mk}^{4} + 8 N^{2} \left(\sum_{m=1}^{M} \gamma_{mk}^{3} \right)  \left(\sum_{n=1}^{M}\gamma_{nk}\right) \right. \\
			& \left. + 2 N^{2} \left(\sum_{m=1}^{M} \gamma_{mk}^{2} \right)^{2} + 4 N^{3} \left(\sum_{m=1}^{M} \gamma_{mk}^{2} \right) \left( \sum_{n=1}^{M}\gamma_{nk} \right)^{2}   \right).
		\end{aligned}
	\end{equation}
	Let $ Y_{u,k} = \rho_{u} \sum\limits_{i\ne k}^{K}B_{k}^{i} + A_{k} +  \rho_{u}C_{k} $, where $ B_{k}^{i} = \abs*{\sum\limits_{m=1}^{M}\mathbf{\hat{g}}_{mk}^{H} \mathbf{\hat{g}}_{mi}}^{2},$ $ A_{k} = \sum\limits_{m=1}^{M}  \norm*{\hat{\mathbf{g}}_{mk}}^{2} $, and $ C_{k} = \sum\limits_{m=1}^{M}\sum\limits_{i=1}^{K}  \left(\beta_{mi} - \gamma_{mi}\right) \norm*{\hat{\mathbf{g}}_{mk}}^{2} $. So, we have
	\begin{equation}\label{Eq:CFOP_YMean}
		\begin{aligned}
			\mathbb{E}\left[Y_{u,k}\right] &= \rho_{u} \sum\limits_{i\ne k}^{K} \mathbb{E}\left[B_{k}^{i} \right] + \mathbb{E}\left[A_{k}\right] +  \rho_{u}\mathbb{E}\left[C_{k} \right], 
		\end{aligned}
	\end{equation}   
	where $  \mathbb{E}\left[B_{k}^{i} \right] = N\sum\limits_{m=1}^{M}\gamma_{mk}\gamma_{mi} + N^{2}\abs*{\sum\limits_{m=1}^{M} \nu_{mk}^{i}}^{2} $, $ \mathbb{E}\left[A_{k}\right] = N\sum\limits_{m=1}^{M}\gamma_{mk} $, $ \mathbb{E}\left[C_{k} \right] = N\sum\limits_{m=1}^{M}\sum\limits_{i=1}^{K}  \left(\beta_{mi} - \gamma_{mi}\right) \gamma_{mk} $  and  $ \nu_{mk}^{i} = c_{mk}c_{mi} \boldsymbol{\phi}_{i}^{H} \mathbf{C}_{\mathbf{y}_{p,m},\mathbf{y}_{p,m}} \boldsymbol{\phi}_{k} $, and $ \mathbf{C}_{\mathbf{y}_{p,m},\mathbf{y}_{p,m}} = \tau_{p}\rho_{p}\sum\limits_{j=1}^{K} \beta_{mj} \boldsymbol{\phi}_{j}\boldsymbol{\phi}_{j}^{H} + \mathbf{I} $. After expanding $ Y_{u,k}^{2} $, we have
	\begin{equation}\label{Eq:CFOP_Y2}
		\begin{aligned}
			\mathbb{E}\left[Y_{u,k}^{2}\right] &=  \rho_{u}^{2} \mathbb{E}\left[\left(\sum_{i\ne k}^{K}B_{k}^{i}\right)^{2}\right] + \mathbb{E}\left[A_{k}^{2}\right] + \rho_{u}^{2}\mathbb{E}\left[C_{k}^{2}\right] \\&+ 2\rho_{u}\sum_{i\ne k}^{K}\mathbb{E}\left[B_{k}^{i}A_{k}\right] + 2\rho_{u}\mathbb{E}\left[A_{k}C_{k}\right] + 2\rho_{u}^{2}\sum_{i\ne k}^{K}\mathbb{E}\left[B_{k}^{i} C_{k}\right].
		\end{aligned}
	\end{equation}
	The expectation $ \mathbb{E}\left[A_{k}^{2}\right] =  \mathbb{E}\left[X_{u,k}\right]/\rho_{u} $ and the expectation of other terms in (\ref{Eq:CFOP_Y2}) is given as follows.
	\begin{equation}\label{Eq:CFOP_MeanSumBki2}
		\begin{aligned}
			\mathbb{E}\left[\left(\sum_{i\ne k}^{K}B_{k}^{i}\right)^{2}\right] 
			&=\sum_{i\ne k}^{K}\sum_{j\ne k }^{K}\Bigg[N\sum_{m=1}^{M}\gamma_{mk}^{2}\gamma_{mi}\gamma_{mj} + N\sum_{m=1}^{M}\gamma_{mk}^{2}\vert \nu_{mi}^{j} \vert^{2} + 2N\sum_{m=1}^{M}\gamma_{mk}\gamma_{mj}\vert \nu_{mk}^{i} \vert^{2}\\
			&+2 N\mathfrak{Re}\left(\sum_{m=1}^{M}\gamma_{mk} (\nu_{mk}^{i})^{*}\nu_{mk}^{j} \left(\nu_{mi}^{j}\right)^{*}\right) +4 N^{2} \mathfrak{Re}\left(\sum_{m=1}^{M}\sum_{n=1}^{M}\gamma_{mk}\gamma_{mi}\nu_{mk}^{j} (\nu_{nk}^{j})^{*} \right) \\
			& +4 N^{2} \mathfrak{Re}\left( \sum_{m=1}^{M}\sum_{n=1}^{M}\gamma_{mk}\nu_{mk}^{i}\nu_{mi}^{j}(\nu_{nk}^{j})^{*} \right) 
			+2 N^{3} \mathfrak{Re}\left(\sum_{m=1}^{M}\sum_{n=1}^{M}\sum_{u =1 }^{M} \nu_{mk}^{i}\nu_{mk}^{j} \left(\nu_{nk}^{i}\right)^{*}\left(\nu_{uk}^{j}\right)^{*}\right) \\&+ 2 N^{3} \mathfrak{Re}\left(\sum_{m=1}^{M}\sum_{n= 1}^{M}\sum_{u =1 }^{M}  \gamma_{mk}\left(\nu_{mi}^{j}\right)^{*}\left(\nu_{nk}^{i}\right)^{*}\nu_{uk}^{j}\right)
			+ N^{2}\abs*{\sum_{m=1}^{M} \nu_{mk}^{i}\nu_{mk}^{j}}^{2}  + N^{2}\abs*{\sum_{m=1}^{M}\gamma_{mk}\nu_{mi}^{j}}^{2} \Bigg] \\&+ \left(\sum_{i\ne k}^{K}\mathbb{E}\left[B_{k}^{i}\right]\right)^{2}.
		\end{aligned}
	\end{equation}
	\begin{equation}\label{MeanCk2}
		\begin{aligned}
			\mathbb{E}\left[C_{k}^{2}\right] 
			&= N\sum_{m = 1}^{M}\sum\limits_{i=1}^{K}  \left(\beta_{mi} - \gamma_{mi}\right)^{2}\gamma_{mk}^{2} + \left(\mathbb{E}\left[C_{k}\right]\right)^{2}.
		\end{aligned}
	\end{equation}
	\begin{equation}\label{MeanAkBkiSimple1}
		\begin{aligned}
			\mathbb{E}\left[B_{k}^{i}A_{k}\right] &= N\sum_{m = 1}^{M}\gamma_{mk}^{2}\gamma_{mi} + N\sum_{m = 1}^{M}\gamma_{mk} \vert \nu_{mk}^{i}\vert^{2} 
			+ 2 N^{2}\mathfrak{Re}\left(\left(\sum_{m = 1}^{M}\gamma_{mk}\nu_{mk}^{i}\right)\left(\sum_{n=1}^{M}\nu_{nk}^{i}\right)^{*}\right)
			\\&+ \mathbb{E}\left[A_{k}\right]\mathbb{E}\left[B_{k}^{i}\right].
		\end{aligned}
	\end{equation}
	\begin{equation}\label{MeanAkCk}
		\begin{aligned}
			\mathbb{E}\left[A_{k}C_{k}\right] &= N\sum_{m = 1}^{M}\sum\limits_{i=1}^{K}  \left(\beta_{mi} - \gamma_{mi}\right)\gamma_{mk}^{2} + \mathbb{E}\left[A_{k}\right]\mathbb{E}\left[C_{k}\right].
		\end{aligned}
	\end{equation}
	Using the result of $ \mathbb{E}\left[B_{k}^{i}A_{k}\right] $, we have
	\begin{equation}\label{MeanCkBkiSimple1}
		\begin{aligned}
			\mathbb{E}\left[B_{k}^{i}C_{k}\right] &= N\sum_{m = 1}^{M}\sum\limits_{j=1}^{K}  \left(\beta_{mj} - \gamma_{mj}\right)\gamma_{mk} \vert \nu_{mk}^{i}\vert^{2} + N\sum_{m = 1}^{M}\sum\limits_{j=1}^{K}  \left(\beta_{mj} - \gamma_{mj}\right)\gamma_{mk}^{2}\gamma_{mi} 
			\\&+ 2\mathfrak{Re} N^{2}\left(\left(\sum_{m = 1}^{M}\sum\limits_{j=1}^{K}  \left(\beta_{mj} - \gamma_{mj}\right)\gamma_{mk}\nu_{mk}^{i}\right)\left(\sum_{n=1}^{M}\nu_{nk}^{i}\right)^{*}\right) + \mathbb{E}\left[C_{k}\right]\mathbb{E}\left[B_{k}^{i}\right].
		\end{aligned}
	\end{equation}
	The correlation of $ X_{u,k} $ and $ Y_{u,k} $ is given as 
	\begin{equation}\label{Eq:CFOP_CorrXY}
		\begin{aligned}
			\mathbb{E}\left[X_{u,k}Y_{u,k}\right] 
			&= \rho_{u}^{2}\sum_{i\ne k}^{K}\mathbb{E}\left[A_{k}^{2}B_{k}^{i}\right] + \rho_{u}\mathbb{E}\left[A_{k}^{3}\right] + \rho_{u}^{2}\mathbb{E}\left[A_{k}^{2}C_{k}\right],
		\end{aligned} 
	\end{equation}
	where
	\begin{equation}
		\begin{aligned}
			\mathbb{E}\left[A_{k}^{2}B_{k}^{i}\right] &=  \mathbb{E}\left[A_{k}^{2} \right] \mathbb{E}\left[B_{k}^{i} \right] + 4 N\sum_{m=1}^{M}\vert \nu_{mk}^{i}\vert^{2}\gamma_{mk}^{2} + 2 N \sum_{m=1}^{M}\gamma_{mk}^{3}\gamma_{mi} 
			+ 2 N^{2}\sum_{m=1}^{M} \sum_{n=1}^{M} \gamma_{mk} \abs*{\nu_{nk}^{i}}^{2}\gamma_{nk} \\&+ 2 N^{2}\sum_{m=1}^{M}\sum_{n=1}^{M} \gamma_{mk}  \gamma_{nk}^{2}\gamma_{ni}
			+ 2 N^{2}\abs*{\sum_{m=1}^{M} \gamma_{mk} \nu_{mk}^{i}}^{2}  + 4 N^{2}\mathfrak{Re}\left( \sum_{m=1}^{M}\sum_{n=1}^{M}\gamma_{mk}^{2} \nu_{mk}^{i} \left(\nu_{nk}^{i}\right)^{*} \right) \\&+ 4N^{2}\mathfrak{Re}\left( \sum_{m=1}^{M}\sum_{n=1}^{M} \sum_{p =1}^{M} \gamma_{mk} \nu_{mk}^{i}  \gamma_{nk}   \left( \nu_{pk}^{i} \right)^{*}  \right).
		\end{aligned}
	\end{equation}
	\begin{equation}\label{MeanAk3}
		\begin{aligned}
			\mathbb{E}\left[A_{k}^{3}\right]
			&= 2N\sum_{m = 1}^{M}\gamma_{mk}^{3} + 3N^{2}\left(\sum_{m = 1}^{M}\gamma_{mk}^{2}\right)\left( \sum_{m = 1}^{M}\gamma_{mk}\right) + N^{3}\left(\sum_{m = 1}^{M}\gamma_{mk}\right)^{3} .
		\end{aligned}
	\end{equation}
	\begin{equation}\label{MeanAk2Ck}
		\begin{aligned}
			\mathbb{E}\left[A_{k}^{2}C_{k}\right] &=  2N\sum_{m = 1}^{M}\sum\limits_{i=1}^{K}  \left(\beta_{mi} - \gamma_{mi}\right)\gamma_{mk}^{3} +  2N^{2}\left(\sum_{m=1}^{M}\sum\limits_{i=1}^{K}  \left(\beta_{mi} - \gamma_{mi}\right)\gamma_{mk}^{2}\right)\left(\sum_{n = 1}^{M}\gamma_{nk}\right)\\& +\mathbb{E}\left[A_{k}^{2}\right]\mathbb{E}\left[C_{k}\right]  .
		\end{aligned}
	\end{equation}
	\section{Proof for Theorem \ref{Thm: CFOP_RateLogNormal}}\label{App: CFOP_ProofRateLogNormal}
	The ergodic rate of $ k $th user is given by
	\begin{equation}
		\begin{aligned}
			R_{u,k}	&= \mathbb{E}\left[\log_{2}\left(1 + \lambda_{u,k}\right)\right].
		\end{aligned}
	\end{equation}
	Since, $ \log_{2}\left(1 + \lambda_{u,k}\right) $ is positive RV, so we have $ R_{u,k}	= \int_{0}^{\infty} \mathbb{P}\left[\log_{2}\left(1 + \lambda_{u,k}\right) > t \right] dt $. 
	The logarithm is a monotonically increasing function. Hence, we have $ R_{u,k}	= \int_{0}^{\infty} \mathbb{P}\left[  \lambda_{u,k} > 2^t - 1 \right] dt $.
	Given $ \lambda_{u,k} \sim \operatorname{LN}\left(\mu_{\lambda_{u,k}},\sigma_{\lambda_{u,k}}^{2}\right) $, the $ R_{u,k} $ is nothing but
	\begin{equation}
		\begin{aligned}
			R_{u,k}	&= \frac{1}{2} \int_{0}^{\infty} \operatorname{erfc}\left(\frac{\ln\left(2^{t} - 1\right) - \mu_{\lambda_{u,k}}}{\sigma_{\lambda_{u,k}} \sqrt{2}}\right) dt.
		\end{aligned}
	\end{equation}
	Using the transformation of variable $ \ln\left(2^{t} - 1\right) = x $, we have
	\begin{equation}
		\begin{aligned}
			R_{u,k} &= \frac{1}{2 \ln 2} \int_{-\infty}^{\infty} \operatorname{erfc}\left(\frac{x - \mu_{\lambda_{u,k}}}{\sigma_{\lambda_{u,k}} \sqrt{2}}\right) \frac{1}{1 + \operatorname{e}^{-x}} dx.
		\end{aligned}
	\end{equation}
	Again, apply $ \frac{x - \mu_{\lambda_{u,k}}}{\sigma_{\lambda_{u,k}} \sqrt{2}} = y $, we have
	\begin{equation}
		\begin{aligned}
			R_{u,k} &= \frac{\sigma_{\lambda_{u,k}}}{\sqrt{2} \ln 2} \int_{-\infty}^{\infty}  \frac{\operatorname{erfc}\left(y\right)}{1 + e^{-\mu_{\lambda_{u,k}} - \sqrt{2} \sigma_{\lambda_{u,k}} y  }} dy, \\
			&=	\frac{\sigma_{\lambda_{u,k}}}{\sqrt{2} \ln 2} \left(\underbrace{ \int_{-\infty}^{0}  \frac{\operatorname{erfc}\left(y\right)}{1 + e^{-\mu_{\lambda_{u,k}} - \sqrt{2} \sigma_{\lambda_{u,k}} y  }} dy}_{I_{1}} + 		\underbrace{ \int_{0}^{\infty}  \frac{\operatorname{erfc}\left(y\right)}{1 + e^{-\mu_{\lambda_{u,k}} - \sqrt{2} \sigma_{\lambda_{u,k}} y  }} dy}_{I_{2}} \right).
		\end{aligned}
	\end{equation}
	\begin{equation}
		\begin{aligned}
			I_{1} &=  \int_{-\infty}^{0}  \frac{\operatorname{erfc}\left(y\right)}{1 + e^{-\mu_{\lambda_{u,k}} - \sqrt{2} \sigma_{\lambda_{u,k}} y  }} dy, 
			=\int_{0}^{\infty}  \frac{2 - \operatorname{erfc}\left(y\right)}{1 + e^{-\mu_{\lambda_{u,k}} + \sqrt{2} \sigma_{\lambda_{u,k}} y  } } dy.
		\end{aligned} 
	\end{equation}
	\begin{equation}
		\begin{aligned}
			I_{1}	&= \underbrace{\int_{0}^{\infty} \frac{2}{1 + e^{-\mu_{\lambda_{u,k}} + \sqrt{2} \sigma_{\lambda_{u,k}} y}} dy }_{I_{1,1}} - \underbrace{\int_{0}^{\infty} \frac{ \operatorname{erfc}\left(y\right)}{1 + e^{-\mu_{\lambda_{u,k}} + \sqrt{2} \sigma_{\lambda_{u,k}} y  } } dy }_{I_{1,2}}.
		\end{aligned}
	\end{equation}
	Hence, we have 
	\begin{equation}
		\begin{aligned}
			R_{u,k} &= \frac{\sigma_{\lambda_{u,k}}}{\sqrt{2} \ln 2}\left( I_{1,1} + I_{2} - I_{1,2} \right).
		\end{aligned}
	\end{equation}
	The integral $I_{1,1}$ is easy to calculate in closed form as follows
	\begin{equation}
		\begin{aligned}
			I_{1,1}	&= \int_{0}^{\infty} \frac{2}{1 + e^{-\mu_{\lambda_{u,k}} + \sqrt{2} \sigma_{\lambda_{u,k}} y}} dy.
		\end{aligned}
	\end{equation}
	Let $ 1 + e^{-\mu_{\lambda_{u,k}} + \sqrt{2} \sigma_{\lambda_{u,k}} y} = u $ then $ dy = \frac{du}{\sqrt{2} \sigma_{\lambda_{u,k}} \left(u-1\right) } $, So we have
	\begin{equation}
		\begin{aligned}
			I_{1,1}	&= \int_{1 + e^{-\mu_{\lambda_{u,k}}}}^{\infty} \frac{2}{u \sqrt{2} \sigma_{\lambda_{u,k}} \left(u-1\right)} du, \\
			&= \frac{\sqrt{2}}{\sigma_{\lambda_{u,k}}} \int_{1 + e^{-\mu_{\lambda_{u,k}}}}^{\infty} \frac{1}{u\left(u-1\right)} du = \frac{\sqrt{2}}{\sigma_{\lambda_{u,k}}} \int_{1 + e^{-\mu_{\lambda_{u,k}}}}^{\infty} \left( \frac{1}{\left(u-1\right)} - \frac{1}{u}\right) du, \\
			&= \frac{\sqrt{2}}{\sigma_{\lambda_{u,k}}} \left(\ln\left(u-1\right) - \ln u\right)\vert_{1 + e^{-\mu_{\lambda_{u,k}}}}^{\infty} = \frac{\sqrt{2}}{\sigma_{\lambda_{u,k}}}  \ln\left(\operatorname{e}^{\mu_{\lambda_{u,k}}} + 1\right).
		\end{aligned}
	\end{equation}
	Now, it is interesting to note that
	\begin{equation}
		\begin{aligned}
			I_{2} - I_{1,2}	&= \int_{0}^{\infty}  \frac{\operatorname{erfc}\left(y\right)}{1 + e^{-\mu_{\lambda_{u,k}} - \sqrt{2} \sigma_{\lambda_{u,k}} y  }}  - \frac{ \operatorname{erfc}\left(y\right)}{1 + e^{-\mu_{\lambda_{u,k}} + \sqrt{2} \sigma_{\lambda_{u,k}} y  } }dy, \\
			&=  \int_{0}^{\infty} \operatorname{erfc}\left(y\right) \left[\frac{\operatorname{e}^{-\mu_{\lambda_{u,k}}} \left(\operatorname{e}^{\sqrt{2} \sigma_{\lambda_{u,k}} y} - \operatorname{e}^{-\sqrt{2} \sigma_{\lambda_{u,k}} y}\right) }{1 + e^{-\mu_{\lambda_{u,k}} - \sqrt{2} \sigma_{\lambda_{u,k}} y  } + e^{-\mu_{\lambda_{u,k}} + \sqrt{2} \sigma_{\lambda_{u,k}} y  } +  e^{-2\mu_{\lambda_{u,k}}} }\right], \\
			&< \frac{\operatorname{e}^{-\mu_{\lambda_{u,k}}}}{1 + \operatorname{e}^{-2\mu_{\lambda_{u,k}}}}\int_{0}^{\infty} \operatorname{erfc}\left(y\right) \left(\operatorname{e}^{\sqrt{2} \sigma_{\lambda_{u,k}} y} - \operatorname{e}^{-\sqrt{2} \sigma_{\lambda_{u,k}} y}\right) dy, \\
			&= \frac{\sqrt{2}\operatorname{e}^{-\mu_{\lambda_{u,k}}}}{\left(1 + \operatorname{e}^{-2\mu_{\lambda_{u,k}}}\right) \sigma_{\lambda_{u,k}} }\left(\operatorname{e}^{\frac{\sigma_{\lambda_{u,k}}^{2}}{2}} - 1\right).
		\end{aligned}
	\end{equation}
	It can be easily shown that $ I_{2} - I_{1,2} > 0 $, so we get a simple upper and lower bound on ergodic rate as follows
	\begin{equation}
		\begin{aligned}
			\frac{\ln\left(\operatorname{e}^{\mu_{\lambda_{u,k}}} + 1\right)}{\ln 2}	< &R_{u,k} <  \frac{1}{\ln 2} \left(  \ln\left(\operatorname{e}^{\mu_{\lambda_{u,k}}} + 1\right) +  \frac{\operatorname{e}^{-\mu_{\lambda_{u,k}}}}{\left(1 + \operatorname{e}^{-2\mu_{\lambda_{u,k}}}\right) }\left(\operatorname{e}^{\frac{\sigma_{\lambda_{u,k}}^{2}}{2}} - 1\right) \right), \\
			\log_{2}\left(\operatorname{e}^{\mu_{\lambda_{u,k}}} + 1\right) < &R_{u,k} < \log_{2}\left(\operatorname{e}^{\mu_{\lambda_{u,k}}} + 1\right) + \frac{\operatorname{e}^{-\mu_{\lambda_{u,k}}}}{\ln 2\left(1 + \operatorname{e}^{-2\mu_{\lambda_{u,k}}}\right) }\left(\operatorname{e}^{\frac{\sigma_{\lambda_{u,k}}^{2}}{2}} - 1\right).
		\end{aligned}
	\end{equation}
	This completes the proof.
	\section{Proof for Lemma \ref{Lem: CFOP_OPUnivariate}}\label{App: CFOP_ProofUnivariateLemma}
	The OP of the $k$th user is given by
	\begin{equation}
		\begin{aligned}
			P_{out}^{k}\left( T \right) = \mathbb{P}\left(\lambda_{u,k} < T \right)= \mathbb{P}\left(X_{u,k} < T Y_{u,k} \right).
		\end{aligned}
	\end{equation}
	Substituting from (\ref{Eq: CFOP_SINRInner}), we have 
	\begin{equation}
		\begin{aligned}
			P_{out}^{k}\left( T \right) &= \mathbb{P}\left(\rho_{u} \norm{\hat{\mathbf{g}}_{k}}^{4} < T \rho_{u} \sum\limits_{\substack{i=1 \\ i\ne k} }^{K} \abs*{\mathbf{\hat{g}}_{k}^{H} \mathbf{\hat{g}}_{i}}^{2}  +   T \hat{\mathbf{g}}_{k}^{H}\left( \rho_{u}\sum\limits_{i=1}^{K} \boldsymbol{\Lambda}_{i} + \mathbf{I}_{MN} \right)\hat{\mathbf{g}}_{k}  \right).
		\end{aligned}
	\end{equation}    
	To simplify the above expression, we first calculate the conditional probability $ P_{out}^{k} $ for a fixed $ \hat{\mathbf{g}}_{k} $. Therefore, for $ \hat{\mathbf{g}}_{k} = \mathbf{b} = \left[\mathbf{b}_{1},\dots,\mathbf{b}_{M} \right]^{T} $, the OP is
	\begin{equation}
		\begin{aligned}
			P_{out}^{k}\left( T \right) \vert \left(\hat{\mathbf{g}}_{k} = \mathbf{b}\right) &= \mathbb{P}\left(\rho_{u} \norm{\mathbf{b}}^{4} < T \rho_{u} \sum\limits_{\substack{i=1 \\ i\ne k} }^{K}\abs*{\mathbf{b}^{H} \hat{\mathbf{g}}_{i}}^{2} \ + \ T \mathbf{b}^{H} \left( \rho_{u}\sum\limits_{i=1}^{K} \boldsymbol{\Lambda}_{i} + \mathbf{I}_{MN} \right) \mathbf{b} \right).
		\end{aligned}
	\end{equation}
	Rearranging the constants to one side of equality, we have
	\begin{equation}\label{Eq:CFOP_CCDFPout}
		\begin{aligned}
			P_{out}^{k}\left( T \right) &=  \mathbb{P}\left(  \sum\limits_{\substack{i=1 \\ i\ne k} }^{K}\vert \mathbf{b}^{H} \hat{\mathbf{g}}_{i} \vert^{2} > d_{k}^{T} \right),
		\end{aligned}
	\end{equation}
	where 
	\begin{equation}\label{Eq:CFOP_dkt}
		\begin{aligned}
			d_{k}^{T} &= \frac{\left(\rho_{u} \left(\sum\limits_{m=1}^{M} \norm*{\mathbf{b}_{m}}^{2}\right)^{2} -  T \left( \rho_{u}\sum\limits_{i=1}^{K}  \sum\limits_{m=1}^{M}\left(\beta_{mi} - \gamma_{mi} \right)\norm*{\mathbf{b}_{m}}^{2} + \sum\limits_{m=1}^{M}\norm*{\mathbf{b}_{m}}^{2} \right) \right)}{T \rho_{u}}.
		\end{aligned}
	\end{equation}
	To further simplify, the CCDF of $\sum\limits_{\substack{i=1 \\ i\ne k} }^{K}\vert \mathbf{b}^{H} \hat{\mathbf{g}}_{i} \vert^{2}$ is required. Hence, note that for the case of no pilot contamination $ Z_{i} = \mathbf{b}^{H} \hat{\mathbf{g}}_{i} $, is a complex Gaussian RV with mean \cite[Eq. 15.25]{kay1993fundamentals} 
	\begin{equation}\label{Eq:CFOP_meanZiNoPilot}
		\begin{aligned}
			\mathbb{E}\left[Z_{i} \vert \hat{\mathbf{g}}_{k} = \mathbf{b} \right]
			&= \mathbf{b}^{H} \mathbb{E}\left[\hat{\mathbf{g}}_{i} \vert \hat{\mathbf{g}}_{k} =\mathbf{b} \right] = 0,
		\end{aligned}
	\end{equation}
	and variance 
	\begin{equation}\label{Eq:CFOP_varianceZiNoPilot}
		\begin{aligned}
			\mathbb{V}\left(Z_{i}\vert \hat{\mathbf{g}}_{k} = \mathbf{b} \right) &= \mathbf{b}^{H} \operatorname{Cov}\left(\hat{\mathbf{g}}_{i} \vert \hat{\mathbf{g}}_{k} \right) \mathbf{b} 
			= \sum_{m=1}^M \norm*{\mathbf{b}_m}^2 \gamma_{mi} := \alpha_{i}. 
		\end{aligned}
	\end{equation}
	Eq. (\ref{Eq:CFOP_meanZiNoPilot}) and (\ref{Eq:CFOP_varianceZiNoPilot}) follow from the fact that $ \hat{\mathbf{g}}_{k} $ and $ \hat{\mathbf{g}}_{i} $ are independent  $\forall i\neq k$. Further, $ \vert Z_{i} \vert^{2} $ is an exponential RV with parameter $\alpha_{i}$.
	Hence, $W = \sum\limits_{\substack{i=1 \\ i\ne k} }^{K}\vert \mathbf{b}^{H} \hat{\mathbf{g}}_{i} \vert^{2}$ is a sum of exponential RVs with CCDF 
	\begin{equation}\label{Eq:CFOP_PoutGivengK}
		\begin{aligned}
			P_{out}^{k} \left(T \right) \vert \left(\hat{\mathbf{g}}_{k} = \mathbf{b}\right)     &= 1 -  \mathbb{P}\left(  W \le d_{k}^{T} \right)  \\
			&= 1 -  \left( \sum\limits_{\substack{i=1 \\ i\ne k} }^{K}  \frac{\alpha_{i}^{K-2}}{\prod\limits^{K}_{{\substack{j=1 \\ j \ne i,j\ne k}}} \left(\alpha_{i} - \alpha_{j} \right)} \left[ 1 -e^{-\frac{d_{k}^{T}}{\alpha_{i}}}  \right] \right)\operatorname{U}\left(d_{k}^{T}\right).
		\end{aligned}
	\end{equation}
	Now that we have obtained the conditional OP, to obtain the final OP, we integrate over the multivariate Gaussian PDF of $ \hat{\mathbf{g}}_{k}$. Hence,
	\begin{equation}\label{Eq:CFOP_PoutOrtho}
		\begin{aligned}
			P_{out}^{k} \left(T \right) &= \int\dots\int \left(1 -  \left( \sum\limits_{\substack{i=1 \\ i\ne k} }^{K}  \frac{\alpha_{i}^{K-2}}{\prod\limits^{K}_{{\substack{j=1 \\ j \ne i, j\ne k}}} \left(\alpha_{i} - \alpha_{j} \right)} \left[ 1 -e^{-\frac{ d_{k}^{T}}{\alpha_{i}}}   \right] \right)\operatorname{U}\left( d_{k}^{T}\right) \right) f_{  \hat{\mathbf{g}}_{k}} \left(\mathbf{b} \right) d \mathbf{b}, \\
			&= 1 - \int\dots\int  \left( \sum\limits_{\substack{i=1 \\ i\ne k} }^{K}  \frac{\alpha_{i}^{K-2}}{\prod\limits^{K}_{{\substack{j=1 \\ j \ne i, j \ne k}}} \left(\alpha_{i} - \alpha_{j} \right)} \left[ 1 -e^{-\frac{ d_{k}^{T}}{\alpha_{i}}}  \right] \right)\operatorname{U}\left(d_{k}^{T}\right)  \prod\limits_{m=1}^{M}\frac{e^{-\frac{\norm*{ \mathbf{b}_{m}}^{2}}{\gamma_{mk}}}}{\pi^{N} \gamma_{mk}^{N} } d \mathbf{b}_{1} \dots d \mathbf{b}_{M} ,    
		\end{aligned}
	\end{equation}
	where the PDF of $\hat{\mathbf{g}}_{k} \left(\mathbf{b}\right)$ is $ f_{\hat{\mathbf{g}}_{k}} \left(\mathbf{b} \right) = \prod\limits_{m=1}^{M}\frac{e^{-\frac{\norm*{\mathbf{b}_{m}}^{2}}{\gamma_{mk}}}}{\pi^{N} \gamma_{mk}^{N} }. $
	After a cartesian to polar transformation, \textit{i.e.}, $b_{mn} = r_{mn}e^{j\phi_{mn}}$, and then using the transformation $\frac{r_{mn}^{2}}{\gamma_{mk}} = x_{mn}$, we get the result in \eqref{Eq: CFOP_PoutOrthoSim2} and this completes the proof.
	\section{Proof for Theorem \ref{Thm: CFOP_OPUnivariate}}\label{App: CFOP_ProofUnivariateTheorem}
	Let $ \mathbf{X} = \left[x_{11},\dots,x_{MN} \right]$ is a random vector with i.i.d. exponential RVs of scale parameter $1$. Then, (\ref{Eq: CFOP_PoutOrthoSim2}) can be described as
	\begin{equation}\label{Eq:CFOP_PoutOrthoSim3}
		\begin{aligned}
			P_{out}^{k} \left(T \right) &= 1 - \mathbb{E}\left[\left( \sum\limits_{\substack{i=1 \\ i\ne k} }^{K}  \frac{\theta_{i}^{K-2}}{\prod\limits^{K}_{{\substack{j=1 \\ j \ne i, j \ne k}}} \left(\theta_{i} - \theta_{j} \right)} \left[ 1 -e^{-\frac{\delta_{k}^{T}}{\theta_{i}}}  \right] \right)\operatorname{U}\left(\delta_{k}^{T}\right) \right], \\
			&= 1 - \int_{\mathbb{R}^{MN}}g\left(\mathbf{x}\right)f_{\mathbf{X}}\left(\mathbf{x}\right) d\mathbf{x}.
		\end{aligned}
	\end{equation}
	Using \cite[eq. 20]{Rahman2004:Integral_DimensionReduction}, \eqref{Eq:CFOP_PoutOrthoSim3} is approximated as
	\begin{equation}\label{Eq: CFOP_Pout_NoPilot_Sim1}
		\begin{aligned}
			P_{out}^{k} \left(T \right) &\approx   1 - \sum_{m=1}^{M}\sum_{n=1}^{N}\mathbb{E}\left[g\left( \mu_{11},\dots,x_{mn},\dots,\mu_{MN}\right) \right] + (MN-1)g\left(\mu_{11},\dots,\mu_{MN}\right),
		\end{aligned}
	\end{equation}
	where $\mu_{mn} = \mathbb{E}[x_{mn}] = 1, \ \forall \ m = 1,\dots,M $ and $n = 1,\dots,N $. Using the values of $ \mu_{mn} $, The $ g\left( \mu_{11},\dots,x_{mn},\dots,\mu_{MN}\right) $ is calculated as
	\begin{equation}\label{Eq: CFOP_gxm}
		\begin{aligned}
			\sum\limits_{\substack{i=1 \\ i\ne k} }^{K} \left(\frac{\left(x_{mn} C_{1,m}^{i} + C_{2,m}^{i}\right)^{K-2}}{\prod\limits^{K}_{{\substack{j=1 \\ j \ne i, j \ne k}}} x_{mn} C_{3,m}^{i,j} + C_{4,m}^{i,j}} \left[ 1 - e^{-\left(\frac{x_{mn}^{2} C_{5,m} + x_{mn} C_{9,m} + C_{10,m}}{ x_{nm} C_{1,m}^{i} + C_{2,m}^{i}}\right)}  \right]\right)\operatorname{U}\left(x_{mn}^{2} C_{5,m} + x_{mn} C_{9,m} + C_{10,m}\right)
		\end{aligned}
	\end{equation} 
	for $ 1 \le m \le M $ and $n = 1,\dots,N $, where $ C_{1,m}^{i} = \gamma_{mk}\gamma_{mi}  $, $ C_{2,m}^{i} = N\sum\limits_{m^{\prime}\ne m}^{M}\gamma_{m^{\prime} k}\gamma_{m^{\prime} i} + \left(N-1\right)\gamma_{mk}\gamma_{mi}$, $ C_{3,m}^{i,j} = C_{1,m}^{i} - C_{1,m}^{j} $, $ C_{4,m}^{i,j} = C_{2,m}^{i} - C_{2,m}^{j} $, $ C_{5,m} = \frac{1}{T}\gamma_{mk}^{2} $,  $C_{6,m} =  \left(N-1\right)\gamma_{mk} +  N \left(\sum\limits_{m^{\prime}\ne m}^{M}\gamma_{m^{\prime} k} \right) $, $C_{7,m} =  \left(\rho_{u}\sum\limits_{i=1}^{K}\left(\beta_{mi} - \gamma_{mi} \right) + 1 \right) \gamma_{mk}$, $C_{8,m} =  N  \sum\limits_{m^{\prime} \ne m }^{M} \left(\rho_{u}\sum\limits_{i=1}^{K}\left(\beta_{m^{\prime}i} - \gamma_{m^{\prime} i} \right) + 1 \right) \gamma_{m^{\prime}k} + \left(N -1\right) C_{7,m}  $, $ C_{9,m} = \frac{2}{T}C_{6,m}\gamma_{mk} - \frac{1}{\rho_{u}} C_{7,m} $, $ C_{10,m} = \frac{ C_{6,m}^{2} }{T} - \frac{1}{\rho_{u}} C_{8,m} $, 
	and 
	\begin{equation}\label{Eq: CFOP_gone}
		\begin{aligned}
			g\left(\mu_{11},\dots,\mu_{MN}\right) &= \left( \sum\limits_{\substack{i=1 \\ i\ne k} }^{K}  \frac{C_{i}^{K-2}}{\prod\limits^{K}_{{\substack{j=1 \\ j \ne i,j\ne k}}} \left(C_{i} - C_{j} \right)} \left[ 1 -e^{-\frac{C_{k}^{T}}{C_{i}}}  \right] \right)\operatorname{U}\left(C_{k}^{T}\right),
		\end{aligned}
	\end{equation}
	where $ C_{i} = N\sum\limits_{m=1}^{M}\gamma_{mk} \gamma_{mi} $ and $ C_{k}^{T} = \frac{N^{2}}{T}\left( \sum\limits_{m=1}^{M} \gamma_{mk}\right)^{2} -  \frac{N}{\rho_{u}} \sum\limits_{m=1}^{M}\gamma_{mk} \left(\rho_{u}\sum\limits_{i=1}^{K}\left(\beta_{mi} - \gamma_{mi} \right) + 1 \right)  $. Finally, the approximation in (\ref{Eq: CFOP_Pout_NoPilot_AppFinal}) is obtained after substituting values from (\ref{Eq: CFOP_gxm}), (\ref{Eq: CFOP_gone}) into (\ref{Eq: CFOP_Pout_NoPilot_Sim1}), and this completes the proof.
	\section{Proof for Corollary \ref{Cor: CFOP_Pout_NPC_mMIMO} }\label{App: CFOP_UnivariatemMIMOCor}
	Note that, for the case of mMIMO, $\gamma_{mk} = \gamma_{k} \ \forall m$ and $\beta_{mk} = \beta_{k}  \ \forall m$. Hence, the \eqref{Eq: CFOP_Pout_NoPilot_AppFinal} simplifies to 
		\begin{equation}\label{Eq: CFOP_Pout_NoPilot_mMIMO}
			\begin{aligned}
				P_{out}^{k} \left(T \right) &\approx   1 - MN \sum\limits_{\substack{i=1 \\ i\ne k} }^{K} \underbrace{\int_{0}^{\infty} \left(D_{1}^{i}\left[ 1 - e^{-\left(D_{2}^{i}x 
						+ D_{3}^{i}\right) }  \right]\right)\operatorname{U}\left(x^{2} D_{4} + x D_{5} + D_{6}\right) e^{-x} dx}_{I}  
				\\ &+ (MN -1) \sum\limits_{\substack{i=1 \\ i\ne k} }^{K}  D_{1}^{i} \left[ 1 -e^{-\left( D_{2}^{i} + D_{3}^{i} \right)}  \right]\operatorname{U}\left(D_{4} + D_{5} + D_{6}\right),
			\end{aligned}
		\end{equation}
	where, 
	\begin{equation}\label{Eq: D1D2D3mMIMO}
		\begin{aligned}
			D_{1}^{i} = \frac{\gamma_{i}^{K-2}}{\prod\limits^{K}_{{\substack{j=1 \\ j \ne i, j\ne k}}}\left(\gamma_{i} - \gamma_{j} \right)}, D_{2}^{i} = \frac{\gamma_{k}}{T \gamma_{i} },   & \quad  
			D_{3}^{i} = \frac{\left( MN-1 \right)\gamma_{k}}{T \gamma_{i}} - \frac{\left( \rho_{u} \sum\limits_{i=1}^{K}\left(\beta_{i} - \gamma_{i} \right) + 1 \right)}{\rho_{u} \gamma_{i}} ,
		\end{aligned}
	\end{equation}
	and
	\begin{equation}\label{Eq: D4D5D6mMIMO}
		\begin{aligned}
			&D_{4} = \frac{1}{T}\gamma_{k}^{2},  \quad 
			D_{5} = \frac{2}{T} \left( MN-1\right)\gamma_{k}^{2}- \frac{1}{\rho_{u}} \gamma_{k} \left( \rho_{u} \sum\limits_{i=1}^{K}\left(\beta_{i} - \gamma_{i} \right) + 1 \right), \\
			&D_{6} = \left( MN-1 \right)\gamma_{k}\left( \frac{\left( MN-1\right)\gamma_{k}}{T } - \frac{\left( \rho_{u} \sum\limits_{i=1}^{K}\left(\beta_{i} - \gamma_{i} \right) + 1 \right)}{\rho_{u}}\right).
		\end{aligned}
	\end{equation} 
	The presence of the unit-step function in Eq. (\ref{Eq: CFOP_Pout_NoPilot_mMIMO}) results in different domains of integration depending on the nature of the roots of a quadratic equation $ D_{4} x_{m}^{2}  + D_{5} x_{m}  + D_{6}=0$. One can easily verify that $D_{5}^{2} - 4D_{4}D_{6} = \frac{\gamma_{k}^{2}}{\rho_{u}^{2}}\left( \rho_{u} \sum\limits_{i=1}^{K}\left(\beta_{i} - \gamma_{i} \right) + 1\right)^{2} > 0$. 
	\subsection{When both the roots are non-positive}\label{App:OutageRootNegativemMIMO}
	In such a scenario, the region of integration will be the entire $\mathbb{R}^{+}$. This is true for $D_{6} \ge 0 $, \textit{i.e.},
	\begin{equation}
		\begin{aligned}
			T \le \frac{\rho_{u} \left( MN-1 \right)\gamma_{k}}{\left( \rho_{u} \sum\limits_{i=1}^{K}\left(\beta_{i} - \gamma_{i} \right) + 1\right)}.
		\end{aligned}
	\end{equation}
	Therefore $I$ reduces to  
	\begin{equation}\label{Eq: IForCase1}
		\begin{aligned}
			I &= D_{1}^{i}\int_{0}^{\infty} \left(\left[ 1 - e^{-\left(D_{2}^{i}x + D_{3}^{i}\right) }  \right]\right)e^{-x} dx = D_{1}^{i} \left[1 -  \frac{e^{-D_{3}^{i}}}{D_{2}^{i} + 1}\right].
		\end{aligned}
	\end{equation}
	\subsection{One root is negative, and the other is positive}
	In this case, the quadratic $D_{4} x^{2}  + D_{5} x  + D_{6} < 0$  for $ 0 \le x \le \kappa $, where 
	\begin{equation}\label{Eq: kappamMIMO}
		\begin{aligned}
			\kappa = \frac{-D_{5} + \sqrt{D_{5}^{2} - 4D_{4}D_{6}}}{2D_{4}}
		\end{aligned}
	\end{equation}
	is the positive root of the quadratic. This is true when $ D_{6} < 0 $, \textit{i.e.},
	\begin{equation}
		\begin{aligned}
			T >  \frac{\rho_{u}\left( MN-1 \right)\gamma_{k}}{\left( \rho_{u} \sum\limits_{i=1}^{K}\left(\beta_{i} - \gamma_{i} \right) + 1\right)}.
		\end{aligned}
	\end{equation}
	Therefore $I$ reduces to  
	\begin{equation}\label{Eq: IForCase2}
		\begin{aligned}
			I &= D_{1}^{i}\int_{\kappa}^{\infty} \left(\left[ 1 - e^{-\left(D_{2}^{i}x 
				+ D_{3}^{i}\right) }  \right]\right)e^{-x} dx = D_{1}^{i} \left[ e^{-\kappa}  - \frac{e^{-D_{3}^{i}} e^{-\kappa (D_{2}^{i} + 1)}}{D_{2}^{i} + 1}\right].
		\end{aligned}
	\end{equation}
	Substitution of \eqref{Eq: IForCase1} and \eqref{Eq: IForCase2} in \eqref{Eq: CFOP_Pout_NoPilot_mMIMO} gives the result in \eqref{Eq: CFOP_Pout_NPCmMIMO_Case1} and \eqref{Eq: CFOP_Pout_NPCmMIMO_Case2}. This completes the proof.
	\bibliographystyle{IEEEtran}
	\bibliography{CFOP_Reference}
\end{document}